\pdfoutput=1
\documentclass[a4paper,oneside]{article}
\usepackage[a4paper]{geometry}
\usepackage[T1]{fontenc}
\usepackage{graphicx}
\usepackage{amsmath}
\usepackage{amssymb}
\usepackage{amsthm}
\usepackage{thmtools,thm-restate}
\usepackage{mathabx,mathrsfs}
\usepackage{mathtools}
\usepackage{bbm}
\usepackage[colorlinks=true, citecolor=blue, urlcolor=blue, pagebackref=true, linktocpage, plainpages=false, pdfpagelabels]{hyperref}

\usepackage{xcolor}
\usepackage[normalem]{ulem}
\usepackage{cancel}
\usepackage{stmaryrd}
\usepackage{subcaption}
\usepackage{paralist}
\usepackage{booktabs}
\usepackage{enumitem}
\usepackage{cleveref}
\usepackage{tikz-cd}
\usepackage{tikz}
\usepackage{undertilde}

\usepackage[textwidth=2.5cm]{todonotes}

\newtheorem{theorem}{Theorem}[section]
\numberwithin{equation}{section}
\newtheorem{lemma}[theorem]{Lemma}

\newtheorem{corollary}[theorem]{Corollary}
\newtheorem{proposition}[theorem]{Proposition}

\theoremstyle{definition}

\newtheorem{remark}[theorem]{Remark}

\newtheorem{example}[theorem]{Example}

\newcommand{\ov}{\overline}

\newcommand{\bbC}{\mathbb{C}}
\newcommand{\bbF}{\mathbb{F}}

\newcommand{\bbN}{\mathbb{N}}

\newcommand{\bbR}{\mathbb{R}}
\newcommand{\bbZ}{\mathbb{Z}}

\newcommand{\calX}{\mathcal{X}}

\newcommand{\specMM}{\Delta_{\mathrm{MM}}}

\newcommand{\degen}{\trianglelefteq}
\newcommand{\neged}{\trianglerighteq}

\newcommand{\sym}{\mathrm{sym}}

\newcommand{\ang}[1]{\langle #1 \rangle}
\newcommand{\Ang}[1]{\left\langle #1 \right\rangle}

\DeclarePairedDelimiter{\abs}{\lvert}{\rvert}

\DeclareMathOperator{\Span}{span}

\DeclareMathOperator{\Rk}{R}
\DeclareMathOperator{\BR}{\underline{R}}
\DeclareMathOperator{\AR}{\utilde{\mathrm{R}}}
\DeclareMathOperator{\AQ}{\utilde{\mathrm{Q}}}

\DeclareMathOperator{\supp}{supp}

\newcommand{\Str}{\mathrm{Str}}
\newcommand{\CW}{\mathrm{CW}}

\newcommand{\mquote}[1]{\text{``}#1\text{''}}

\definecolor{diffdelete}{RGB}{180,35,35}
\definecolor{diffinsert}{RGB}{20,85,170}

\newcommand{\email}[1]{\href{mailto:#1}{\texttt{#1}}}

\title{Semidirect sums of tensors \\ and the matrix multiplication exponent}
\author{Baitian Li\thanks{Columbia University. \email{bl3052@columbia.edu}.}}
\date{}

\begin{document}

\maketitle

\begin{abstract}
  This note is an exposition of the recent preprints from OpenAI that establish the bound $\omega \leq 9/4$ on the matrix multiplication exponent.

  We argue that the heart of these preprints is a new technique, which we formulate as the \emph{untwisting lemma} for \emph{semidirect sums} of tensors. This gives a generally applicable method for analyzing tensors. We also discuss further consequences of this method, as well as the relation between the OpenAI proof and previous work.
\end{abstract}

\section{Introduction} \label{sec:intro}

Determining the exponent $\omega$ of matrix multiplication is one of the foundational questions in algebraic complexity theory. It has been known since Strassen~\cite{Str69Gaussian} that determining $\omega$ amounts to studying the asymptotic rank of matrix multiplication tensors. Let $\Rk(T)$ and $\BR(T)$ denote the rank and the border rank of a tensor $T$. The \emph{asymptotic rank} of $T$ is defined by
\[ \AR(T) \coloneqq \lim_{n\to\infty} \Rk(T^{\otimes n})^{1/n}. \]
Then $\omega = \log_2 \AR(\ang{2,2,2})$, and more generally $\omega = \log_m \AR(\ang{m,m,m})$ for any integer $m \geq 2$.

The long quest for the matrix multiplication exponent can be divided into two kinds of efforts: developing new frameworks for bounding the exponent, and discovering instantiations that are effective within such frameworks.

In his foundational work~\cite{Str69Gaussian}, Strassen observed that any finite estimate $\Rk(\ang{n,n,n}) \leq r$ gives the bound $n^\omega \leq r$; together with his discovery that $\Rk(\ang{2,2,2}) \leq 7$, this yields the first nontrivial bound $\omega \leq \log_2 7$. It was then observed~\cite{BCLR79BRk,Bini80BRk} that border rank bounds for rectangular matrix multiplication also suffice: if $\BR(\ang{E,H,L}) \leq r$, then $(EHL)^{\omega/3} \leq r$.

The next milestone is Sch\"onhage's celebrated \emph{asymptotic sum inequality}~\cite{Sch81Tau}, which later became a cornerstone of the study of asymptotic rank. It shows that one can even use border rank bounds for direct sums of matrix multiplication tensors to bound $\omega$:
\begin{center}
  if \; $\displaystyle \BR\left(\bigoplus_{i=1}^q \ang{E_i, H_i, L_i}\right) \leq r$, \quad
  then \; $\displaystyle \sum_{i=1}^q (E_i H_i L_i)^{\omega / 3} \leq r$.
\end{center}
Sch\"onhage complemented this beautiful inequality with his discovery of the \emph{direct sum identity}
\[ \BR\bigl(\ang{n, 1, m} \oplus \ang{1, (n-1)(m-1), 1}\bigr) = nm + 1. \]
This is very surprising: the first summand alone already has border rank $nm$, yet adding the second summand costs only one extra unit, whereas computing it separately would require $(n-1)(m-1)$.

The next conceptual breakthrough again came from Strassen, who introduced the \emph{laser method}~\cite{Str87RelBilin}. It makes it possible to obtain bounds from tensors that are sums of matrix multiplication tensors which are \emph{not} direct sums. The method was then refined by Coppersmith and Winograd~\cite{CW90Laser} to make it more widely applicable, and they applied it to the family of tensors now known as the big Coppersmith--Winograd tensors $\CW_q$. All subsequent improvements over more than three decades~\cite{Sto10Thesis,DS13MM,Wil12MM,LeGall14MM,AW21Refined,DWZ23Hashing,VXXZ24Rect,ADWXXZ25MoreAsym,DEKMRSZAWB26AlphaEvolve} rely on increasingly refined analyses of modified versions of the laser method applied to $\CW_q$. The best published bound along this line of work is $\omega < 2.371177$~\cite{DEKMRSZAWB26AlphaEvolve}, which improves upon the bound $\omega < 2.3755$ of Coppersmith and Winograd only in the third decimal place. Moreover, as shown by Ambainis, Filmus and Le Gall~\cite{AFL15Limit}, a broad generalization of the laser method applied to $\CW_q$ cannot yield any bound better than $\omega < 2.3078$.

Very recently, OpenAI released three preprints~\cite{OpenAI26Staggered,OpenAI26Below2258,OpenAI26NineFourths} that culminate in the substantial improvement $\omega \leq 9/4 = 2.25$. In this exposition, we give a streamlined proof of this result, identify the key technical breakthrough behind it, and discuss its connections with previous work.

To explain these connections, we first go back to the time when Strassen invented the laser method~\cite{Str87RelBilin}.

\paragraph{Notation.} We follow the notation that has become standard in the matrix multiplication literature over the past decade; we list some of it here. For a detailed treatment of the preliminaries, we refer the reader to~\cite[Chap.~15]{BCS97ACT} or~\cite{Blaser13FMM}. A tensor $T$ lives in a tensor space $U \otimes V \otimes W$, and the spaces $U, V, W$ are called the three \emph{modes} of the tensor. Basis vectors of $U$, $V$, $W$ are written as $x_i$, $y_j$, $z_k$, respectively. We write $\leq$ and $\degen$ for restriction and degeneration of tensors, and $\lesssim$ (or $\gtrsim$) for asymptotic restriction. The diagonal tensor is $\ang{n} = \sum_{i=1}^n x_i y_i z_i$, and the matrix multiplication tensor is $\ang{E, H, L} = \sum_{i=1}^E \sum_{j=1}^H \sum_{k=1}^L x_{ij} y_{jk} z_{ki}$.

\subsection{The semidirect sum of tensors} \label{subsec:semidirect-sum}

Let $T_1, T_2$ be two tensors whose third modes are the same distinguished space $W$, that is, $T_1 \in U_1 \otimes V_1 \otimes W$ and $T_2 \in U_2 \otimes V_2 \otimes W$. We call their sum $T_1 + T_2$, viewed in the ambient space $(U_1 \oplus U_2) \otimes (V_1 \oplus V_2) \otimes W$, the \emph{semidirect sum} of $T_1$ and $T_2$, and denote it by $T_1 \boxplus_3 T_2$; the subscript $3$ indicates that the two tensors share the third mode. When the shared mode is clear from context, we usually drop the subscript.

\paragraph{The Strassen tensor.}
Semidirect sums of tensors appear frequently in the study of matrix multiplication, although they have not been formally defined. In the paper where Strassen introduced the \emph{laser method}~\cite{Str87RelBilin}, the tensor he analyzed is the following one, which he called the null algebra and which is now known as the Strassen tensor:
\[ \Str_q = \underbracket{\sum_{i=1}^q x_0 y_i z_i}_{T_1} + \underbracket{\sum_{i=1}^q x_i y_0 z_i}_{T_2}. \]
This is naturally a semidirect sum.

The tensor $\Str_q$ has minimal border rank $q+1$, as the following border decomposition shows:
\[ \Str_q = \lim_{\lambda \to 0} \frac{1}{\lambda} \left( \sum_{i=1}^q (x_0 + \lambda x_i)(y_0 + \lambda y_i)z_i - x_0 y_0 \sum_{i=1}^q z_i \right). \]

An interesting feature of the Strassen tensor is that both of its semidirect summands $T_1, T_2$ and its ``outer structure'' are matrix multiplication tensors. More precisely, each of them is an inner product:
\begin{itemize}
  \item The first summand $T_1$ is an inner product on the $V$ and $W$ modes, which we write as $\ang{1, 1, q}$.
  \item The second summand $T_2$ is an inner product on the $U$ and $W$ modes, which we write as $\ang{q, 1, 1}$.
  \item Writing $T_1 \in U_1 \otimes V_1 \otimes W$ and $T_2 \in U_2 \otimes V_2 \otimes W$, the \emph{blockwise support} of $\Str_q$ is
  \[ \Str_q \; \in \; U_1 \otimes V_1 \otimes W \; \oplus \; U_2 \otimes V_2 \otimes W, \]
  which has the shape of the tensor $u_1 v_1 w + u_2 v_2 w = \ang{1, 2, 1}$.
\end{itemize}

\paragraph{The ``schizophrenia''.}
One might hope to exploit this by swapping the roles of some of the $x$- and $y$-variables:
\[ \Str_q = \sum_{i=1}^q (x_0 y_i + y_0 x_i) z_i. \]
With the variables ordered as above, this looks just like the matrix multiplication tensor $\ang{1, 2, q}$:
\[ \sum_{i=1}^q (x_0 y_i + y_0 x_i) z_i  \quad  \mquote{\cong} \quad \sum_{i=1}^q \sum_{j=1}^2 (x'_{1j}y'_{ji})z'_{i1} = \ang{1, 2, q}. \]
If this argument were valid, we would obtain something like $\BR(\ang{1, 2, q}) \leq q+1$, which would in turn give
\begin{equation}
  \omega \leq 3\cdot \frac{\log(q+1)} {\log(2q)}. \label{eqn:desired-estimate}
\end{equation}
Taking $q=3$, this would give $\omega < 2.322$.

However, the argument is plainly false: the tensor $\ang{1, 2, q}$ is known to have border rank exactly $2q$. The flaw is that permuting variables across different modes destroys the (border) rank decomposition.

What if we return to the semidirect sum formalism instead? We first write $\Str_q$ as
\[ \Str_q = \ang{1, 1, q} \boxplus \ang{q, 1, 1}, \]
and take its tensor square. By the ``binomial theorem'', we would obtain
\begin{align*}
  \Str_q^{\otimes 2} &= \ang{1, 1, q}^{\otimes 2} \boxplus (\ang{1, 1, q} \otimes \ang{q, 1, 1})^{\boxplus 2} \boxplus \ang{q, 1, 1}^{\otimes 2}\\
  &= \ang{1, 1, q^2} \boxplus \ang{q, 1, q}^{\boxplus 2} \boxplus \ang{q^2, 1, 1}.
\end{align*}
More generally, the $2n$-th tensor power $\Str_q^{\otimes 2n}$ should contain $\binom{2n}{n}$ semidirect summands of the form $\ang{q^n, 1, q^n}$. This suggests the restriction
\begin{equation}
  \Str_q^{\otimes 2n} \geq \Ang{q^n, \binom{2n}{n}, q^n}. \label{eqn:hypothetical-restriction}
\end{equation}
If this were true, we could use $(q^{2n} \cdot \binom{2n}{n})^{\omega/3} \leq (q+1)^{2n}$, which again yields the desired estimate \eqref{eqn:desired-estimate}. The above heuristic leads one to conjecture an ``AM--GM type inequality'' for semidirect sums of tensors:
\[ \ang{q, 1, 1} \boxplus \ang{1, 1, q} \gtrsim \mquote{\ang{\sqrt q, 2, \sqrt q}}, \]
or, more ambitiously, for general tensors, provided a suitable notion of ``square root'' can be defined:
\[ T_1 \boxplus T_2 \gtrsim \mquote{\ang{1, 2, 1} \otimes \sqrt{T_1} \otimes \sqrt{T_2}}. \]

However, this argument already breaks down at the very first step. Let us expand the square of the Strassen tensor and examine it carefully:
\[ \Str_q^{\otimes 2} = \sum_{i=1}^q\sum_{j=1}^q x_{00}y_{ij}z_{ij}
+ \sum_{i=1}^q\sum_{j=1}^q x_{0j}y_{i0}z_{ij}
+ \sum_{i=1}^q\sum_{j=1}^q x_{i0}y_{0j}z_{ij}
+ \sum_{i=1}^q\sum_{j=1}^q x_{ij}y_{00}z_{ij}. \]
The ``central binomial'' part consists of the two middle terms:
\[ \Str_q^{\otimes 2} \geq \sum_{i=1}^q\sum_{j=1}^q x_{0j}y_{i0}z_{ij}
+ \sum_{i=1}^q\sum_{j=1}^q x_{i0}y_{0j}z_{ij}. \]

In their paper generalizing Strassen's laser method, Coppersmith and Winograd~\cite{CW90Laser} reflected on exactly this feature of the Strassen tensor. They described the indices of $z$ as \emph{``schizophrenic''}: in the first summand, $i$ is indexed by $y$, so it behaves like a column index, while in the second summand, $i$ is indexed by $x$, so it behaves like a row index. Consequently, these two matrix multiplication tensors cannot be combined into a single larger one!

Let us look at this again through the lens of semidirect sums. The issue is that we wishfully identified $T_1 \otimes T_2 \boxplus T_2 \otimes T_1$ with $T_1 \otimes T_2 \boxplus T_1 \otimes T_2$. Locally, we certainly have $T_1 \otimes T_2 \cong T_2 \otimes T_1$, by permuting the tensor factors within each mode. The problem is that this isomorphism also permutes the factors of the shared third mode, and the semidirect sum is sensitive to exactly this.

For this reason, Strassen did not arrive at the estimate \eqref{eqn:desired-estimate}. Instead, he introduced the laser method to alleviate this ``schizophrenia''.

\paragraph{The laser method.}
One first takes the $n$-th power $\Str_q^{\otimes n}$ and then tensors it with its two cyclic permutations. The resulting tensor has outer structure $\ang{2^n, 2^n, 2^n}$, and each block contains a matrix multiplication tensor of volume $q^{3n}$. That is, the resulting tensor can be written as
\[ T = \sum_{i,j,k=1}^{2^n} T_{ijk}, \qquad T_{ijk} \in U_{ij} \otimes V_{jk} \otimes W_{ki}, \]
with respect to direct sum decompositions $U = \bigoplus U_{ij}$, $V = \bigoplus V_{jk}$ and $W = \bigoplus W_{ki}$ of its three modes. The laser method~\cite{Str87RelBilin} shows that one can monomially degenerate the outer structure so that what remains is a direct sum of about $4^n$ genuine matrix multiplication tensors. Indeed, fix a target integer $\ell$, and let $\lambda$ be a formal parameter. Since
\[ (i+j+k-\ell)^2 = \underbracket{(i-\ell)^2 + 2(i-\ell)j}_{U_{ij}} + \underbracket{j^2 + 2jk}_{V_{jk}} + \underbracket{k^2 + 2k(i-\ell)}_{W_{ki}}, \]
scaling each block $U_{ij}$, $V_{jk}$, $W_{ki}$ by $\lambda$ to the power of the corresponding underbraced term shows that
\[ T \cong \sum_{i,j,k=1}^{2^n} \lambda^{(i+j+k-\ell)^2} \, T_{ijk}. \]
As $(i+j+k-\ell)^2 \geq 0$, with equality if and only if $i+j+k = \ell$, letting $\lambda \to 0$ gives the degeneration
\[ T \neged \bigoplus_{i+j+k = \ell} T_{ijk}. \]
The right-hand side is indeed a direct sum: if $i+j+k = \ell$, then any two of $i,j,k$ determine the third, so distinct summands use disjoint blocks in each mode. Choosing $\ell = 3 \cdot 2^{n-1} + 1$, the number of summands is exactly $3 \cdot 4^{n-1}$.
The asymptotic sum inequality then gives
\[ 3\cdot 4^{n-1} \cdot q^{3n \cdot \omega/3} \leq (q+1)^{3n}, \]
and letting $n \to \infty$ yields $\omega \leq \log_q \bigl((q+1)^3 / 4\bigr)$. Strassen's bound $\omega < 2.479$ follows by plugging in $q=5$.

In other words, instead of letting the outer structure $\ang{2^n, 2^n, 2^n}$ genuinely play the role of a matrix multiplication tensor, which would give
\begin{align*}
  \AR( \ang{2,2,2} \otimes \ang{q,q,q} ) \leq (q+1)^3 &\implies 2^{\omega} \cdot q^{\omega} \leq (q+1)^{3},
  \intertext{the laser method sacrifices the outer structure and changes the game to}
  \AR( \ang{2\cdot 2} \otimes \ang{q,q,q} ) \leq (q+1)^3 &\implies 2^{2} \cdot q^{\omega} \leq (q+1)^{3}.
\end{align*}
Since $\omega \geq 2$, the latter can never give a better bound.

The later development of the laser method can be seen as a continued fight against this ``schizophrenia''. In their celebrated work, Coppersmith and Winograd~\cite{CW90Laser} proposed a strengthening of Strassen's laser method, in which the outer structure can be zeroed out to a direct sum even when it is not of matrix multiplication shape. Applying their method to their big tensor $\CW_q$, the ``first-order analysis'' gives $\omega < 2.388$. They further improved the exponent to $\omega < 2.376$ by analyzing the square $\CW_q^{\otimes 2}$. This is counterintuitive at first glance: both analyses study tensor powers of a fixed tensor, and $\CW_q$ and $\CW_q^{\otimes 2}$ have essentially the same family of tensor powers. The subtlety is that, in the analysis of $\CW_q^{\otimes 2}$, they chose a combinatorial outer structure different from the naive square of the first-order one, under which some matrix multiplication tensors merge into larger ones; this advantage is invisible to the first-order analysis. All subsequent improvements~\cite{Sto10Thesis,DS13MM,Wil12MM,LeGall14MM,AW21Refined,DWZ23Hashing,VXXZ24Rect,ADWXXZ25MoreAsym,DEKMRSZAWB26AlphaEvolve} turn out to push this gain further, while balancing it against certain losses in the hierarchical structure of tensor powers.

Nevertheless, whether the ``schizophrenia'' can be completely ``cured'' has remained an elusive goal. Even though the naive argument for \eqref{eqn:hypothetical-restriction} above is false, could the desired restriction nonetheless hold, at least asymptotically? If so, plugging $q=3$ into \eqref{eqn:desired-estimate} would give $\omega < 2.322$, surpassing the best previous bound, which is based on analyzing $\CW_5$.

The recent preprints from OpenAI~\cite{OpenAI26Below2258,OpenAI26NineFourths} show that the answer is \emph{yes}.

\section{The untwisting lemma} \label{sec:untwisting}

In this section we explain the ``treatment'' of the ``schizophrenia'' of tensors. For tensors $T$ and $S$, we say that $T$ \emph{asymptotically restricts} to $S$, and write $T \gtrsim S$, if $\ang{f_n} \otimes T^{\otimes n} \geq S^{\otimes n}$ for some $f_n = 2^{o(n)}$.

\begin{theorem}[Untwisting lemma] \label{thm:semidirect-sum-restriction}
  Let $T_1, \dots, T_q$ be tensors with $T_i \in U_i \otimes V_i \otimes W$, each isomorphic to a fixed tensor $T$, where the isomorphisms are allowed to act nontrivially on the shared third space $W$. Then
  \[ \bigboxplus_{i=1}^q T_i \gtrsim \ang{1, q, 1} \otimes T. \]
\end{theorem}

Although the above lemma requires the tensors $T_i$ to be isomorphic to each other, a standard argument similar to Sch\"onhage's proof of the asymptotic sum inequality~\cite{Sch81Tau} boosts it to a version in which the $T_i$ need not be isomorphic. Since we have not yet introduced the language needed to state such a version, we defer it to \Cref{subsec:functional}.

\begin{example}
  As a first illustration, the lemma immediately shows that the hypothetical restriction \eqref{eqn:hypothetical-restriction} we dreamed of earlier does hold \emph{asymptotically}. Recall that the $2n$-th tensor power of $\Str_q = \ang{q, 1, 1} \boxplus \ang{1, 1, q}$ is a semidirect sum of $2^{2n}$ matrix multiplication tensors, exactly $\binom{2n}{n}$ of which are isomorphic to $\ang{q^n, 1, q^n}$. Restricting to these $\binom{2n}{n}$ summands and applying \Cref{thm:semidirect-sum-restriction}, we obtain
  \[ \Str_q^{\otimes 2n} \gtrsim \Ang{1, \binom{2n}{n}, 1} \otimes \ang{q^n, 1, q^n} = \Ang{q^n, \binom{2n}{n}, q^n}, \]
  which is the asymptotic version of \eqref{eqn:hypothetical-restriction}. In other words, a single application of \Cref{thm:semidirect-sum-restriction} to the Strassen tensor already improves Strassen's laser method bound $\omega < 2.479$ (obtained with $q = 5$) to $\omega < 2.322$ (obtained with $q = 3$)!
\end{example}

Let us now remark on where this lemma comes from. Recently, OpenAI released three preprints improving the upper bound on $\omega$. Chronologically, the first one~\cite{OpenAI26Staggered} improves the previous bound only slightly, via a technically involved higher-order analysis of the laser method.

The second preprint~\cite{OpenAI26Below2258} proves the substantially stronger bound $\omega < 2.258$, and its Lemma~3.3 essentially implies \Cref{thm:semidirect-sum-restriction}. The main test object of the second preprint is the big Coppersmith--Winograd tensor $\CW_1$, which is in fact the structure tensor of the truncated polynomial algebra $\bbF[x]/(x^3)$.

In the third preprint~\cite{OpenAI26NineFourths}, the test objects are replaced by the family of polynomial multiplication tensors $\bbF[x]_{<n} \times \bbF[x]_{<m} \to \bbF[x]_{<n+m-1}$, without truncation, which yields the bound $\omega \leq 9/4$. We also remark that Proposition~3.1 of the third preprint proves another technical statement which at first glance looks different from Lemma~3.3 of the second preprint; nevertheless, both lead to the same desired result, which we formulate as \Cref{thm:semidirect-sum-restriction}.

We now present the proof of \Cref{thm:semidirect-sum-restriction}, for which we first need a lemma. Let $\ang{\bbF[t] / (t^p)}$ denote the structure tensor of the truncated polynomial algebra, namely
\[ \ang{\bbF[t] / (t^p)} = \sum_{\substack{j, k \geq 0 \\ j + k < p}} x_j y_k z_{j + k}. \]

\begin{lemma} \label{lem:semidirect-sum-degeneration}
  Let $T_1,\dots,T_q$ be tensors with $T_i \in U_i \otimes V_i \otimes W$, which need not be isomorphic. If $p \geq 2q$, then we have the degeneration
  \[ \ang{\bbF[t] / (t^p)} \otimes \bigboxplus_{i=1}^q T_i \; \neged \; \ang{1, q, 1} \otimes \bigoplus_{i=1}^q T_i. \]
\end{lemma}

For readers familiar with the laser method, we remark that the following proof is in fact an ``asymmetric'' application of Strassen's laser method. Strassen's original laser method~\cite[\S7]{Str87RelBilin} deals with a \emph{balanced} tight structure, meaning that the numbers of occupied outer blocks in the three modes are the same. Given $N$ blocks in each mode, it shows how to monomially degenerate the structure to $\Omega(N)$ block triples, no two of which share a block in any mode. The asymmetric version starts with an outer block structure with $N$, $N$ and $M$ blocks in the three modes, where $N \gg M$, and monomially degenerates it to $\Omega(N)$ block triples, any two of which can share a block only in the third mode. This type of asymmetric argument is known as \emph{asymmetric hashing} in the literature~\cite[\S9]{CW90Laser}, \cite{DWZ23Hashing}.

\begin{proof}
  We first decompose the three mode spaces of the left-hand side into blocks. The first mode is $\mathcal U = \bbF^{p} \otimes (U_1 \oplus \cdots \oplus U_q)$; expanding $\bbF^p \otimes U_i = U_i^{\oplus p}$, we write $\mathcal U$ as the direct sum of distinguished copies $U_i^{(j)}$ of $U_i$, where $1 \leq i \leq q$ and $0 \leq j < p$. Similarly, the second mode is the direct sum of copies $V_i^{(k)}$, and the third mode is the direct sum of copies $W^{(\ell)}$, where $0 \leq k, \ell < p$. With respect to this block structure, the left-hand side is the sum of distinguished copies $T_i^{(jk)}$ of $T_i$, one for each $i$ and each pair $j, k \geq 0$ with $j + k < p$, where
  \[ T_i^{(jk)} \in U_i^{(j)} \otimes V_i^{(k)} \otimes W^{(j + k)}. \]

  We now construct a monomial degeneration. Write $c_i = i + q - 1$. We multiply the block $T_i^{(jk)}$ by $\lambda^{(j + k - c_i)^2}$, where $\lambda$ is a formal parameter. This is realized by a monomial degeneration, thanks to the factorization
  \[ \lambda^{(j + k - c_i)^2} = \underbracket{\lambda^{c_i^2 - 2 j c_i}}_{U_i^{(j)}} \cdot \underbracket{\lambda^{-2 k c_i}}_{V_i^{(k)}} \cdot \underbracket{\lambda^{\ell^2}}_{W^{(\ell)}}, \qquad \ell = j + k. \]
  Note that the exponent on $W^{(\ell)}$ depends only on $\ell$ and not on $i$, which is essential since the third mode is shared.

  Since $(j + k - c_i)^2$ is a square, it is nonnegative and vanishes only when $j + k = c_i$. Hence, as $\lambda \to 0$, the surviving blocks are exactly those with $j + k = c_i$. This eliminates the variable $k = c_i - j$, so every surviving block $T_i^{(jk)}$ lies in
  \[ U_i^{(j)} \otimes V_i^{(c_i - j)} \otimes W^{(c_i)}, \qquad 0 \leq j \leq c_i. \]
  Let $T_i^\star$ denote the sum of these surviving blocks. Since $p \geq 2q$, we have $c_i \leq 2q - 1 < p$, so all $c_i + 1$ such blocks are present. Therefore:
  \begin{enumerate}
    \item The outer structure of $T_i^\star$ is a $(c_i + 1)$-term inner product, hence $T_i^\star \cong \ang{1, q + i, 1} \otimes T_i \geq \ang{1, q, 1} \otimes T_i$.
    \item The tensors $T_i^\star$ occupy pairwise disjoint subspaces, since $T_i^\star \in U_i^{\oplus p} \otimes V_i^{\oplus p} \otimes W^{(c_i)}$ and the $c_i$ are pairwise distinct.
  \end{enumerate}
  Consequently,
  \[ \ang{\bbF[t] / (t^p)} \otimes \bigboxplus_{i=1}^q T_i \; \neged \; \bigoplus_{i=1}^q T_i^\star \; \geq \; \bigoplus_{i=1}^q \ang{1, q, 1} \otimes T_i. \qedhere \]
\end{proof}

We now make several remarks on the above proof and its relation to the OpenAI preprints.

\begin{remark}
  The third OpenAI preprint essentially carries out the argument of \Cref{lem:semidirect-sum-degeneration} in~\cite[Prop.~3.1]{OpenAI26NineFourths}, whose statement directly combines it with a (border) rank upper bound for $\ang{\bbF[t]/(t^p)}$. We keep these two ingredients separate and use the border rank bound only later, in the proof of \Cref{thm:semidirect-sum-restriction}, so that our lemma isolates the purely combinatorial part of the argument. By comparison, the second preprint~\cite[Lem.~3.3]{OpenAI26Below2258} uses the zeroing-out method, which can be seen as the Coppersmith--Winograd version of the above argument.
\end{remark}

\begin{remark}
  Historically, the laser method of Strassen~\cite{Str87RelBilin} and that of Coppersmith and Winograd~\cite{CW90Laser} look superficially different: the former is a monomial degeneration and retains a constant fraction of the coordinates, while the latter is a zeroing out and retains only a subpolynomial fraction of them. They are nevertheless intrinsically related; we give a thorough discussion of this connection in \Cref{app:two-laser-methods}.
\end{remark}

\begin{proof}[Proof of \Cref{thm:semidirect-sum-restriction}]
  Take $p = 2q$ in \Cref{lem:semidirect-sum-degeneration}. Since $\ang{\bbF[t]/(t^p)}$ has minimal border rank $p$, i.e., $\ang{p} \neged \ang{\bbF[t]/(t^p)}$,\footnote{For instance, a uniform construction is to take $a_k = \lambda^k$: the algebra $\bbF[t]/\bigl(\prod_{k=1}^p (t - a_k)\bigr)$ is isomorphic to $\bbF^p$ for $\lambda \neq 0$, and it degenerates to $\bbF[t]/(t^p)$ as $\lambda \to 0$.} we obtain
  \[ \ang{2q} \otimes \bigboxplus_{i=1}^q T_i \; \neged \; \ang{\bbF[t]/(t^{2q})} \otimes \bigboxplus_{i=1}^q T_i \; \neged \; \ang{1,q,1} \otimes \bigoplus_{i=1}^q T_i. \]
  Since every $T_i$ is isomorphic to $T$, the right-hand side is isomorphic to $\ang{1,q,1} \otimes \ang{q} \otimes T$, so
  \[ \ang{2q} \otimes \bigboxplus_{i=1}^q T_i \; \neged \; \ang{q} \otimes \ang{1,q,1} \otimes T. \]
  Viewing $\ang{2q}$ as $\ang{2} \otimes \ang{q}$, we can cancel the factor $\ang{q}$ on both sides, at the price of replacing the degeneration by an asymptotic restriction. We remind the reader that this is the \emph{multiplicative cancellation} property, which we prove in \Cref{lem:cancel} below.\footnote{More generally, cancellation still holds when both the hypothesis and the conclusion are asymptotic restrictions; see~\cite[Eqn.~1.11]{Str88AsymSpec} or~\cite[Lem.~3.7]{WZ26Spectra}.} We get
  \[ \ang{2} \otimes \bigboxplus_{i=1}^q T_i \gtrsim \ang{1,q,1} \otimes T. \]
  It seems that we have lost a factor of $2$. However, we can bootstrap the result by the \emph{tensor power trick}. For every $r \geq 1$, the $r$-th tensor power $\bigl(\bigboxplus_{i=1}^q T_i\bigr)^{\otimes r}$ is a semidirect sum of $q^r$ isomorphic copies of $T^{\otimes r}$, sharing the third space $W^{\otimes r}$. Applying what we have just proved to it, we obtain
  \[ \ang{2} \otimes \Bigl(\bigboxplus_{i=1}^q T_i\Bigr)^{\otimes r} \gtrsim \ang{1, q^r, 1} \otimes T^{\otimes r} = (\ang{1, q, 1} \otimes T)^{\otimes r}. \]
  This means that the restriction $\ang{f_n} \otimes \bigl(\bigboxplus_{i=1}^q T_i\bigr)^{\otimes n} \geq (\ang{1, q, 1} \otimes T)^{\otimes n}$ holds with $f_n \leq 2^{n/r + o(n)}$. Letting $r \to \infty$ shows that the desired asymptotic restriction holds.
\end{proof}

\begin{lemma}[Multiplicative cancellation] \label{lem:cancel}
  Let $C$ be a nonzero tensor. If $C \otimes T \neged C \otimes S$, then $T \gtrsim S$.
\end{lemma}
\begin{proof}
  We first prove $C \otimes T^{\otimes n} \neged C \otimes S^{\otimes n}$ by induction on $n$. The case $n = 1$ is the assumption. If it holds for $n-1$, then
  \[ C \otimes T^{\otimes n} = (C \otimes T^{\otimes (n-1)}) \otimes T \neged (C \otimes S^{\otimes (n-1)}) \otimes T = (C \otimes T) \otimes S^{\otimes (n-1)} \neged C \otimes S^{\otimes n}. \]
  Let $r = \BR(C)$, so that $\ang{r} \neged C$. Since $C$ is nonzero, we also have $C \geq \ang{1}$. Therefore
  \[ \ang{r} \otimes T^{\otimes n} \neged C \otimes T^{\otimes n} \neged C \otimes S^{\otimes n} \geq S^{\otimes n}. \]
  This shows that $T$ asymptotically degenerates to $S$, which we know is equivalent to asymptotic restriction.
\end{proof}

\subsection{Functional estimates} \label{subsec:functional}

A first consequence of the untwisting lemma, which can be seen as a natural generalization of the bound $\omega < 2.322$ shown above, is an estimate, which we call \emph{semi-superadditivity}, for two classes of tensor functionals. The first is the \emph{value}, an auxiliary function introduced by Coppersmith and Winograd~\cite{CW90Laser} and used in all later analyses of the laser method. The second comes from the \emph{asymptotic spectrum}, a family of functionals introduced by Strassen~\cite{Str88AsymSpec} to systematically study the asymptotic properties of tensors. In \Cref{sec:poly-mult} we will see that both of them can be used to prove $\omega \leq 9/4$; the OpenAI preprint~\cite{OpenAI26NineFourths} uses the latter.

\paragraph{Values.} For a parameter $\tau \geq 2/3$, the value $V_\tau$ of a tensor $T$ is defined by
\[ V_\tau(T) \coloneqq \sup \left\{ \left(\sum_{i=1}^m (E_i H_i L_i)^\tau\right)^{1/n} : T^{\otimes n} \neged \bigoplus_{i=1}^m \ang{E_i, H_i, L_i} \right\}. \]
The choice of degeneration $\neged$ here is merely a convention; it is not hard to see that replacing it by restriction or by asymptotic restriction yields the same function. It is well known that $V_\tau$ is superadditive and supermultiplicative~\cite[\S8]{CW90Laser}:
\[ V_\tau(T \oplus S) \geq V_\tau(T) + V_\tau(S), \qquad V_\tau(T \otimes S) \geq V_\tau(T) \cdot V_\tau(S). \]
We now show that $V_\tau$ also behaves well under semidirect sums.

\begin{theorem}[Semi-superadditivity of the value] \label{thm:value-semidirect}
  For any semidirect sum $T \boxplus S$, we have
  \[ V_\tau(T \boxplus S)^{1/\tau} \geq V_\tau(T)^{1/\tau} + V_\tau(S)^{1/\tau}. \]
\end{theorem}
\begin{proof}
  The proof is a standard concentration argument. For each $0 \leq k \leq n$, the $n$-th tensor power $(T \boxplus S)^{\otimes n}$ restricts to a semidirect sum of $\binom{n}{k}$ isomorphic copies of $T^{\otimes k} \otimes S^{\otimes (n-k)}$. Hence by \Cref{thm:semidirect-sum-restriction}, together with $V_\tau(\ang{1, C, 1}) \geq C^\tau$ and the supermultiplicativity of $V_\tau$, we obtain
  \[ V_\tau(T \boxplus S)^n = V_\tau\bigl((T \boxplus S)^{\otimes n}\bigr) \geq \binom{n}{k}^{\tau} V_\tau(T)^k \, V_\tau(S)^{n-k}. \]
  Taking $1/\tau$-th powers and summing over $k$, we obtain
  \begin{align*}
    (n+1) \, V_\tau(T \boxplus S)^{n/\tau} &\geq \sum_{k=0}^n \binom{n}{k} V_\tau(T)^{k/\tau} \, V_\tau(S)^{(n-k)/\tau} \\
    &= \bigl(V_\tau(T)^{1/\tau} + V_\tau(S)^{1/\tau}\bigr)^n.
  \end{align*}
  Taking $n$-th roots and letting $n \to \infty$ gives the desired estimate.
\end{proof}

\paragraph{Asymptotic spectrum.}
A similar inequality holds for spectral points in the sense of Strassen's asymptotic spectrum~\cite{Str88AsymSpec,WZ26Spectra}. The \emph{asymptotic spectrum of tensors} $\calX$ is the set of all functionals $\phi$ from tensors to nonnegative reals that are
\begin{itemize}
  \item (monotone) $\phi(S) \leq \phi(T)$ whenever $S \leq T$;
  \item (normalized) $\phi(\ang{1}) = 1$;
  \item (additive) $\phi(S \oplus T) = \phi(S) + \phi(T)$ for all $S, T$;
  \item (multiplicative) $\phi(S \otimes T) = \phi(S) \cdot \phi(T)$ for all $S, T$.
\end{itemize}
Its elements are called \emph{spectral points}. A basic way to study spectral points is to parameterize them by their behavior on matrix multiplication tensors. It is a standard fact~\cite[Prop.~4.3]{Str88AsymSpec} that the restriction of any spectral point $\phi$ to matrix multiplication tensors is determined by three exponents $\theta_1, \theta_2, \theta_3$:
\[ \phi(\ang{E, H, L}) = E^{\theta_1} H^{\theta_2} L^{\theta_3} \quad \text{for all } E, H, L. \]
We call the set of all exponent vectors $(\theta_1, \theta_2, \theta_3)$ arising in this way the (logarithmic) \emph{spectrum of matrix multiplication}, and denote it by $\specMM \subset \bbR^3$.

We have the following analogue of \Cref{thm:value-semidirect}.

\begin{theorem}[Semi-superadditivity of spectral points] \label{thm:spectral-semidirect}
  Let $\phi$ be a spectral point and $\varkappa \in \{1, 2, 3\}$ with $\theta_\varkappa > 0$. Then
  \[ \phi(T \boxplus_{\varkappa + 1} S)^{1/\theta_\varkappa} \geq \phi(T)^{1/\theta_\varkappa} + \phi(S)^{1/\theta_\varkappa}, \]
  where the index $\varkappa + 1$ is taken modulo $3$.
\end{theorem}
The proof is almost identical; the only difference is that one has to keep track of which mode is shared, which determines the shape of the inner product produced by \Cref{thm:semidirect-sum-restriction}. We omit the details.

\subsection{More consequences}

We list some further applications of the untwisting lemma.

\paragraph{Support rank.} Consider a tensor of the form
\[ T = \sum_{i=1}^E \sum_{j=1}^H \sum_{k=1}^L t_{ijk} \cdot x_{ij} y_{jk} z_{ki} \]
with all coefficients $t_{ijk} \neq 0$; in other words, $T$ has the same support as the matrix multiplication tensor $\ang{E, H, L}$. The \emph{support rank} $\Rk_s(\ang{E, H, L})$ is the minimum rank of such a tensor $T$. Cohn and Umans~\cite{CU13Coherent} introduced the corresponding \emph{support rank exponent}
\[ \omega_s \coloneqq \inf \bigl\{ \tau : \Rk_s(\ang{n, n, n}) = O(n^\tau) \bigr\}. \]
Clearly $\omega_s \leq \omega$, and Cohn and Umans proved the partial converse $\omega \leq (3\omega_s - 2)/2$, i.e., $\omega - 2 \leq \frac{3}{2} (\omega_s - 2)$. Using the untwisting lemma, we can easily show that the two exponents in fact coincide.

\begin{lemma} \label{lem:support-restriction}
  If $T$ has the same support as $\ang{E, H, L}$, then $T \gtrsim \ang{E, H, L}$.
\end{lemma}
\begin{proof}
  Grouping the terms of $T$ according to $j$, we write $T = \sum_{j=1}^H T_j$, where
  \[ T_j = \sum_{i=1}^E \sum_{k=1}^L t_{ijk} \cdot x_{ij} y_{jk} z_{ki}. \]
  For distinct $j$, the tensors $T_j$ involve disjoint sets of $x$- and $y$-variables, while all of them share the $z$-variables; hence $T = \bigboxplus_{j=1}^H T_j$ is a semidirect sum. Moreover, rescaling $z_{ki} \mapsto t_{ijk}^{-1} z_{ki}$ shows that each $T_j$ is isomorphic to $\ang{E, 1, L}$. Note that this isomorphism depends on $j$ and acts nontrivially on the shared third mode, which is exactly the situation allowed by \Cref{thm:semidirect-sum-restriction}. Applying it, we obtain $T \gtrsim \ang{1, H, 1} \otimes \ang{E, 1, L} = \ang{E, H, L}$.
\end{proof}

\begin{corollary}
  $\omega_s = \omega$.
\end{corollary}
\begin{proof}
  It suffices to show $\omega \leq \omega_s$. For every $n$, let $T$ be a tensor with the same support as $\ang{n, n, n}$ and $\Rk(T) = \Rk_s(\ang{n, n, n})$. By \Cref{lem:support-restriction}, $T \gtrsim \ang{n, n, n}$, and hence $n^\omega \leq \Rk(T) = \Rk_s(\ang{n, n, n})$. Therefore $\omega \leq \omega_s$.
\end{proof}

\paragraph{Partial matrix multiplication.} We keep the notation above, but now allow some of the coefficients $t_{ijk}$ to vanish; we call such a tensor $T$ a \emph{partial and weighted matrix multiplication tensor}, and write $\supp t \coloneqq \{(i, j, k) : t_{ijk} \neq 0\}$. The argument above generalizes easily to this setting. Such tensors already appeared in the work of Bini, Capovani, Lotti and Romani~\cite{BCLR79BRk}, where border rank was introduced, and were later studied systematically by Sch\"onhage~\cite{Sch81Tau}.

More precisely, Sch\"onhage~\cite[Thm.~4.1]{Sch81Tau} treated the case where $T$ is a \emph{partial} matrix multiplication, i.e., $T$ is obtained from a matrix multiplication tensor by zeroing out some variables in two of its modes. He showed that such a $T$ asymptotically degenerates to a \emph{total} (i.e., genuine) matrix multiplication tensor of volume $\abs{\supp t}$.

The case where $\supp t$ is an arbitrary subset, not necessarily obtained by zeroing out variables, was studied by Vrana~\cite{Vrana24Partial}, who showed that $T$ then asymptotically degenerates to a tensor whose support is that of a matrix multiplication tensor of volume $\abs{\supp t}$. Combined with \Cref{lem:support-restriction}, this shows that $T$ in fact asymptotically degenerates to a genuine matrix multiplication tensor of volume $\abs{\supp t}$. In particular, we obtain the following estimate, for which we give a direct proof via \Cref{thm:value-semidirect}.

\begin{theorem} \label{thm:partial-value}
  Every partial and weighted matrix multiplication tensor $T$ satisfies $V_\tau(T) \geq \abs{\supp t}^\tau$.
\end{theorem}
\begin{proof}
  Suppose $\supp t \subseteq [E] \times [H] \times [L]$. We proceed in three steps, peeling off one index at a time.

  First, suppose $H = L = 1$. Then
  \[ T = \sum_{i=1}^E t_{i11} \cdot x_{i1} y_{11} z_{1i}, \]
  which, after discarding the unused variables and rescaling, is isomorphic to $\ang{\abs{\supp t}, 1, 1}$. Hence $V_\tau(T) = \abs{\supp t}^\tau$.

  Next, suppose only $H = 1$. Grouping the terms according to $k$, we write
  \[ T = \sum_{i=1}^E \sum_{k=1}^L t_{i1k} \cdot x_{i1} y_{1k} z_{ki} = \sum_{k=1}^L T_k, \qquad T_k = \sum_{i=1}^E t_{i1k} \cdot x_{i1} y_{1k} z_{ki}. \]
  The tensors $T_k$ share the $x$-variables and involve disjoint sets of $y$- and $z$-variables, so $T = \bigboxplus_{k=1}^L T_k$ is a semidirect sum sharing the first mode. Each $T_k$ falls into the first case, with support of size $\abs{\supp t \cap ([E] \times \{1\} \times \{k\})}$. Applying \Cref{thm:value-semidirect} repeatedly, we obtain
  \[ V_\tau(T)^{1/\tau} \geq \sum_{k=1}^L V_\tau(T_k)^{1/\tau} = \sum_{k=1}^L \abs{\supp t \cap ([E] \times \{1\} \times \{k\})} = \abs{\supp t}. \]

  Finally, for general $H$, grouping the terms according to $j$ writes $T = \bigboxplus_{j=1}^H T_j$ as a semidirect sum sharing the third mode, as in the proof of \Cref{lem:support-restriction}, where each $T_j$ falls into the second case. The same computation gives $V_\tau(T)^{1/\tau} \geq \sum_{j=1}^H V_\tau(T_j)^{1/\tau} \geq \abs{\supp t}$.
\end{proof}

\subsection{Barriers surpassed by the untwisting lemma}

We now have the key tool for proving the new bound $\omega \leq 9/4$. Before doing so, we reflect on the effectiveness of this tool and on the barriers it may face. All previous methods for bounding the matrix multiplication exponent can be viewed as consisting of two steps:
\begin{enumerate}[label=(\arabic*)]
  \item start with a tensor $T$ with a sufficiently good asymptotic rank estimate, which usually comes directly from a border rank bound;
  \item determine the most valuable (direct sums of) matrix multiplication tensors to which $T^{\otimes n}$ can be restricted.
\end{enumerate}
The technical crux is how much freedom one has in step~(2).

Researchers have proposed several restricted paradigms for step~(2) and studied their barriers. The Ambainis--Filmus--Le Gall barrier~\cite{AFL15Limit} concerns a variant of the laser method. Since the laser method, as well as the group-theoretic approach of Cohn and Umans~\cite{CU03Group}, first fixes a distinguished basis and then analyzes possible zeroings out of the resulting combinatorial structure, Alman and Vassilevska Williams~\cite{AW18Further,AW18Limits} formalized this type of approach as the \emph{solar method}. They also considered the more general \emph{galactic method}, which allows monomial degenerations, and the \emph{universal method}, which allows arbitrary degenerations.

We point out that the new estimates obtained via the untwisting lemma are not subject to the Ambainis--Filmus--Le Gall barrier, nor to the known barriers for the solar and galactic methods, since the argument does not fit into the frameworks in which these barriers are proved. At first glance, the untwisting lemma is proved via the laser method itself, but the technical twist is that it introduces a \emph{catalyst}: one first tensors with the truncated polynomial algebra $\ang{\bbF[t]/(t^p)}$. The degeneration obtained from the laser method is combinatorial, but the upper bound on the border rank of $\ang{\bbF[t]/(t^p)}$ is highly non-combinatorial, as it relies on an explicit border rank decomposition. Moreover, this catalyst is effectively used several times, through multiplicative cancellation (\Cref{lem:cancel}), which completely scrambles the coefficients of the resulting tensor degeneration.

However, since one still seeks degenerations from tensor powers $T^{\otimes n}$ (even though the new proof does not fit entirely into this framework), the approach remains within the scope of the universal method, and is therefore subject to the slice rank barrier~\cite{Alm21Barrier}, also known as the irreversibility barrier~\cite{CVZ21Barrier}. For example, even if the untwisting lemma is incorporated into the analysis of the big Coppersmith--Winograd tensors $\CW_q$, one still cannot prove any bound better than $\omega < 2.16805$.

Roughly speaking, suppose that we start with a tensor $T$ with asymptotic rank $\AR(T) = r$. The irreversibility barrier says that if the asymptotic subrank of $T$ is noticeably smaller, say $\AQ(T) \leq r^{1-\delta}$, then no analysis based on $T$ can prove a bound better than $\omega < 2/(1-\delta)$. This gives strong barriers for the previously used tensors, since their asymptotic subranks are known to drop, with a uniform bound on $\delta$.

In contrast, the new OpenAI preprint~\cite{OpenAI26NineFourths} works with polynomial multiplication tensors. We denote by $P_{p,q}$ the structure tensor of polynomial multiplication $\bbF[t]_{<p} \times \bbF[t]_{<q} \to \bbF[t]_{<p+q-1}$, that is,
\[ P_{p,q} = \sum_{i=0}^{p-1} \sum_{j=0}^{q-1} x_i y_j z_{i+j}. \]
It is known that this family of tensors has minimal rank $p+q-1$, and that in the balanced case $p = q$ the tensor $P_{q,q}$ moreover has linear subrank; therefore $\AQ(P_{q,q}) \geq \Omega(q)$. In other words, although for each fixed $q$ the irreversibility barrier still rules out proving $\omega = 2$ from $P_{q,q}$ alone, the resulting bounds tend to $2$ as $q \to \infty$: for the family $P_{q,q}$ as a whole, we currently know of no barrier towards proving $\omega = 2$! In hindsight, this may explain, at a philosophical level, why the new $\omega \leq 9/4$ result is based on this family of tensors.

Compared with the polynomial multiplication tensor $P_{q,q}$, the tensor $\CW_q$ seems much less valuable, as it is ``hollow'' in comparison. The reason that laser method analyses still use the Coppersmith--Winograd tensors is partly that, within those frameworks, there was no good way to take advantage of the rich middle part of polynomial multiplication tensors. This appears to be no longer the case once semidirect sums are available.

\section{Bounds via polynomial multiplication} \label{sec:poly-mult}

This section is organized as follows. In the OpenAI preprint~\cite{OpenAI26NineFourths}, the two key tensor degenerations are intertwined with the inequalities for spectral points that they imply. In \Cref{subsec:constructions}, we disentangle them and state them as standalone tensor degenerations. In \Cref{subsec:estimating}, we first prove the bound $\omega \leq 9/4$ using values, the classical tool of the laser method literature, and then reproduce the proof using spectral points. Finally, in \Cref{subsec:closer-look}, we take a closer look at the two proofs: beyond the superficial similarity that both reduce to solving the same system of inequalities, we explain the different implications of the bounds for the two types of functionals.

\subsection{Constructions} \label{subsec:constructions}

To construct certain tensor degenerations, it is sometimes more convenient to take the viewpoint of bilinear maps instead of tensors. Recall that a bilinear map $T \colon U \times V \to W$ is the same as a tensor in $U^* \otimes V^* \otimes W$. By abuse of notation, we use the same letter for both, and refer to the modes simply as $U$, $V$ and $W$. Note that under this identification, a subspace of an input space $V$ corresponds to a quotient of the mode $V^*$, and vice versa.

\begin{lemma} \label{lem:filtration-degen}
  Let $T \colon U \times V \to W$ be a bilinear map, and let $V_1 \subseteq V$ and $W_1 \subseteq W$ be subspaces such that $T(U, V_1) \subseteq W_1$. Then $T$ induces two bilinear maps:
  \begin{itemize}
    \item the \emph{subtensor} $T_1 \colon U \times V_1 \to W_1$, and
    \item the \emph{quotient tensor} $T_2 \colon U \times (V/V_1) \to W/W_1$.
  \end{itemize}
  Moreover, we have the degeneration $T \neged T_1 \boxplus_1 T_2$.
\end{lemma}
\begin{proof}
  Choose complements $V = V_1 \oplus V_2$ and $W = W_1 \oplus W_2$. Since $T(U, V_1) \subseteq W_1$, the tensor $T$ has no component in $U \otimes V_1 \otimes W_2$. Scaling $V_1$ by $\lambda^{-1}$ and $W_1$ by $\lambda$, the components in $U \otimes V_1 \otimes W_1$ and $U \otimes V_2 \otimes W_2$ are unchanged, while the component in $U \otimes V_2 \otimes W_1$ is multiplied by $\lambda$. Letting $\lambda \to 0$ gives a monomial degeneration to $T_1 \boxplus_1 T_2$.
\end{proof}

\begin{proposition} \label{prop:P-split}
  For $q \geq 2$, we have
  \[ \ang{1,1,2} \otimes P_{p,q} \neged P_{p,q-1} \boxplus_1 P_{p,q+1}. \]
\end{proposition}
\begin{proof}
  The left-hand side is the bilinear map $T \colon U \times V \to W$ with
  \[ U = \bbF[t]_{<p}, \quad V = \bbF[t]_{<q}^{\oplus 2}, \quad W = \bbF[t]_{<p+q-1}^{\oplus 2}, \]
  given by polynomial multiplication
  \[ \bigl(a, (b_1, b_2)\bigr) \mapsto (a b_1, a b_2), \]
  where $a \in \bbF[t]_{<p}$ and $b_1, b_2 \in \bbF[t]_{<q}$. Consider the $\bbF[t]$-linear map
  \[ \phi \colon \bbF[t]^{\oplus 2} \to \bbF[t], \quad (b_1, b_2) \mapsto b_1 + t b_2, \]
  and take the subspaces
  \[ V_1 = (\ker \phi) \cap V, \quad W_1 = (\ker \phi) \cap W, \]
  that is, the pairs with $b_1 = -t b_2$. Since $\phi$ is $\bbF[t]$-linear, we have $T(U, V_1) \subseteq W_1$. The subtensor $T_1 \colon U \times V_1 \to W_1$ is exactly $P_{p,q-1}$, since $V_1 \cong \bbF[t]_{<q-1}$ and $W_1 \cong \bbF[t]_{<p+q-2}$ via $(-t b, b) \mapsto b$. For the quotient tensor $T_2 \colon U \times (V/V_1) \to W/W_1$, the map $\phi$ induces isomorphisms $V/V_1 \cong \phi(V) = \bbF[t]_{<q+1}$ and $W/W_1 \cong \phi(W) = \bbF[t]_{<p+q}$, under which $T_2$ becomes $P_{p,q+1}$. The claim now follows from \Cref{lem:filtration-degen}.
\end{proof}

Let $Q_{p,q}$ denote the tensor obtained from $P_{p,q}$ by swapping the second and third modes, that is,
\[ Q_{p,q} = \sum_{i=0}^{p-1} \sum_{j=0}^{q-1} x_i y_{i+j} z_j. \]

\begin{proposition} \label{prop:P-Q-split}
  We have
  \[ P_{p,3q+p-1} \neged (\ang{1,1,2} \otimes P_{p,q}) \boxplus_1 Q_{p,q}. \]
\end{proposition}
\begin{proof}
  We split the $y$-variables into three consecutive blocks
  \[ Y_1 = \{y_0, \dots, y_{q-1}\}, \quad Y_2 = \{y_q, \dots, y_{2q+p-2}\}, \quad Y_3 = \{y_{2q+p-1}, \dots, y_{3q+p-2}\}, \]
  of sizes $q$, $q+p-1$ and $q$, respectively. Let $V_1 = \Span(Y_1 \sqcup Y_3)$, and let $W_1$ be the image of $U \times V_1$ under $P_{p,3q+p-1}$. Then $W_1$ is spanned by the $z$-variables
  \[ z_0, \dots, z_{p+q-2} \quad \text{and} \quad z_{2q+p-1}, \dots, z_{3q+2p-3}. \]
  These two segments do not overlap, so the subtensor $T_1 \colon U \times V_1 \to W_1$ simply multiplies a polynomial in $\bbF[t]_{<p}$ with each of two independent polynomials in $\bbF[t]_{<q}$. Hence $T_1 \cong \ang{1,1,2} \otimes P_{p,q}$.

  For the quotient tensor $T_2 \colon U \times (V/V_1) \to W/W_1$, the quotient spaces are spanned by $Y_2$ and by $z_{p+q-1}, \dots, z_{2q+p-2}$, respectively. Thus
  \[ T_2 = \sum_{i=0}^{p-1} \sum_{k=0}^{q-1} x_i \, y_{k + q + p - 1 - i} \, z_{k + p + q - 1}, \]
  and after relabeling $x_i \mapsto x_{p-1-i}$, $y_{j} \mapsto y_{j-q}$ and $z_{\ell} \mapsto z_{\ell-p-q+1}$, this is exactly $Q_{p,q}$. The claim now follows from \Cref{lem:filtration-degen}.
\end{proof}

\subsection{Estimating the exponent} \label{subsec:estimating}

We now have all the tensor degenerations in hand; the remaining task is to analyze how they lead to an estimate of $\omega$. Our overall strategy is still to take advantage of the tensors $P_{p,p}$ of minimal rank $\Rk(P_{p,p}) = 2p-1$, and to eventually trade these intermediate tensors $P_{p,p}$ for matrix multiplication tensors.

For now, let us ignore the twist that some tensors get their modes permuted in \Cref{prop:P-Q-split}, and record the two relations we intuitively have:
\begin{itemize}
  \item (concavity, \Cref{prop:P-split}) $P_{p,q} \to (P_{p,q-1} + P_{p,q+1}) / 2$;
  \item (descent, \Cref{prop:P-Q-split}) $P_{p,3q+p-1} \to 3P_{p,q}$.
\end{itemize}
Intuitively, the smaller $p$ or $q$ is, the closer $P_{p,q}$ is to a matrix multiplication tensor; indeed, when $p = 1$ or $q = 1$, it is exactly one. We would therefore like to gradually reduce $P_{p,q}$ to smaller tensors. A natural goal is to turn $P_{p,q+1}$ into several ``copies'' of $P_{p,q}$, where the copies are counted not in the direct sum sense, but in the matrix multiplication sense. One can first apply concavity once to obtain $P_{p,q}/2$, at the cost of also producing $P_{p,q+2}/2$. One can then apply concavity again at $(p, q+2)$ to turn the latter into $P_{p,q+1}/4$ and $P_{p,q+3}/4$, and so on. This process does not climb indefinitely: as soon as we reach $P_{p,r}$ with $r = 3q+p-1$, descent brings it directly back down to $3P_{p,q}$.

To make this involved procedure rigorous, we can use existing tools that reduce it to simple algebraic inequalities. The OpenAI preprint~\cite{OpenAI26NineFourths} uses the formalism of the asymptotic spectrum. This is not very surprising, as one motivation of Strassen's theory is to better understand the asymptotic sum inequality~\cite{Str88AsymSpec}, and there are known examples of bounds obtained via the asymptotic spectrum~\cite{AL26Speedup}. We point out that one can also use a traditional tool from the laser method literature, namely the notion of \emph{value}, with a slight technical twist. We present both arguments, in \Cref{subsubsec:value-estimates} and \Cref{subsubsec:spectral-estimates} respectively, and then discuss their relation.

\subsubsection{Value estimates} \label{subsubsec:value-estimates}

We define the \emph{catalytic value} by
\[ \ov{V_\tau}(T) \coloneqq \sup \left\{ \left(\frac{E_1 H_1 L_1}{E_0 H_0 L_0} \right)^{\tau/n} : \ang{E_0,H_0,L_0} \otimes T^{\otimes n} \neged \ang{E_1, H_1, L_1} \right\}. \]

Compared with the classical value, there are two technical differences: we allow a catalyst $\ang{E_0,H_0,L_0}$ on the left-hand side, but we no longer allow a direct sum on the right-hand side.

Moreover, the parameter $\tau$ no longer plays any role: the objective $\bigl(E_1 H_1 L_1 / (E_0 H_0 L_0)\bigr)^{\tau/n}$ is the $\tau$-th power of the same expression with $\tau = 1$, so $\ov{V_\tau}(T) = \ov{V_1}(T)^\tau$. From now on we simply write $\ov{V} = \ov{V_1}$, that is,
\[ \ov{V}(T) = \sup \left\{ \left(\frac{E_1 H_1 L_1}{E_0 H_0 L_0} \right)^{1/n} : \ang{E_0,H_0,L_0} \otimes T^{\otimes n} \neged \ang{E_1, H_1, L_1} \right\}. \]

The catalytic value is still supermultiplicative, just like the classical value~\cite[\S8]{CW90Laser}, but it is no longer obviously superadditive. We do not need superadditivity, since none of the tensor degenerations we use produce direct sums. Moreover, the proof of \Cref{thm:value-semidirect} carries over to the catalytic value, and the catalytic value can also be used to bound $\omega$. We collect these properties below.

\begin{proposition} \label{prop:catalytic-value}
  The catalytic value $\ov{V}$ is supermultiplicative and semi-superadditive, that is,
  \[ \ov{V}(T \otimes S) \geq \ov{V}(T) \cdot \ov{V}(S), \qquad \ov{V}(T \boxplus S) \geq \ov{V}(T) + \ov{V}(S), \]
  and $\ov{V}(\ang{1}) = 1$. Moreover, it satisfies the following \emph{$\ang{E,H,L}$-linearity}:
  \[ \ov{V}(\ang{E, H, L} \otimes T) = EHL \cdot \ov{V}(T). \]
  Finally, we have $\ov{V}(T) \leq \AR(T)^{3/\omega}$.
\end{proposition}
\begin{proof}
  Supermultiplicativity is proved by tensoring degenerations, and semi-superadditivity by the same argument as in the proof of \Cref{thm:value-semidirect}; note that the latter only uses the largest term in the binomial expansion, and hence does not use superadditivity. For the normalization, $\ov{V}(\ang{1}) \geq 1$ is trivial. Conversely, if $\ang{E_0,H_0,L_0} \neged \ang{E_1,H_1,L_1}$, then comparing the three flattening ranks gives $E_0 H_0 \geq E_1 H_1$, $H_0 L_0 \geq H_1 L_1$ and $L_0 E_0 \geq L_1 E_1$, whose product yields $E_1 H_1 L_1 \leq E_0 H_0 L_0$. Hence $\ov{V}(\ang{1}) \leq 1$.

  Next, supermultiplicativity gives $\ov{V}(\ang{E,H,L} \otimes T) \geq EHL \cdot \ov{V}(T)$. For the reverse inequality, consider any degeneration
  \begin{equation}
    \ang{E_0,H_0,L_0} \otimes (\ang{E,H,L} \otimes T)^{\otimes n} \neged \ang{E_1, H_1, L_1}. \label{eqn:catalytic-degen}
  \end{equation}
  Its left-hand side equals $\ang{E_0 E^n, H_0 H^n, L_0 L^n} \otimes T^{\otimes n}$, so \eqref{eqn:catalytic-degen} is also a catalytic degeneration for $T$, with the larger catalyst $\ang{E_0 E^n, H_0 H^n, L_0 L^n}$. Comparing the two normalizations, the contribution of \eqref{eqn:catalytic-degen} to $\ov{V}(T)$ is exactly $(EHL)^{-1}$ times its contribution to $\ov{V}(\ang{E,H,L} \otimes T)$. Taking the supremum gives $\ov{V}(\ang{E,H,L} \otimes T) \leq EHL \cdot \ov{V}(T)$.

  Finally, consider any degeneration
  \[ \ang{E_0,H_0,L_0} \otimes T^{\otimes n} \neged \ang{E_1, H_1, L_1}. \]
  Tensoring it with its two cyclic permutations gives
  \[ \ang{N_0,N_0,N_0} \otimes (T \otimes T^{\sigma} \otimes T^{\sigma^2})^{\otimes n} \neged \ang{N_1,N_1,N_1}, \]
  where $N_0 = E_0 H_0 L_0$, $N_1 = E_1 H_1 L_1$, and $\sigma$ denotes the cyclic permutation of the modes. Since $\AR(\ang{N,N,N}) = N^\omega$ and $\AR(T \otimes T^{\sigma} \otimes T^{\sigma^2}) \leq \AR(T)^3$, taking asymptotic ranks gives $N_1^\omega \leq N_0^\omega \AR(T)^{3n}$, that is,
  \[ \left( \frac{E_1 H_1 L_1}{E_0 H_0 L_0} \right)^{1/n} \leq \AR(T)^{3/\omega}. \]
  Taking the supremum over all such degenerations gives the desired bound.
\end{proof}

The $\ang{E,H,L}$-linearity is the only genuinely new property we need. In \Cref{prop:P-split}, we want to estimate the ``value'' of the left-hand side $\ang{1,1,2} \otimes P_{p,q}$. For the classical value we only know $V_\tau(\ang{1,1,2} \otimes P_{p,q}) \geq 2^\tau \, V_\tau(P_{p,q})$, and it is not clear whether equality holds.

Let $\nu_{p,q} = \ov{V}(P_{p,q})$. We now collect the relations satisfied by $\nu$. The catalytic value is invariant under permuting the three modes, since such a permutation maps matrix multiplication tensors to matrix multiplication tensors with the same volume $EHL$. Since $P_{q,p}$ and $Q_{p,q}$ are both obtained from $P_{p,q}$ by permuting the modes, we have
\begin{itemize}
  \item (symmetry) $\nu_{p,q} = \nu_{q,p} = \ov{V}(Q_{p,q})$.
\end{itemize}
By \Cref{prop:catalytic-value}, the constructions in the previous subsection imply the following inequalities:
\begin{itemize}
  \item (\Cref{prop:P-split} $\implies$ concavity) $\nu_{p,q} \geq (\nu_{p,q-1} + \nu_{p,q+1}) / 2$ for $q \geq 2$;
  \item (\Cref{prop:P-Q-split} $\implies$ descent) $\nu_{p,3q+p-1} \geq 3\nu_{p,q}$.
\end{itemize}
Finally, $P_{p,1} \cong \ang{p,1,1}$ is itself a matrix multiplication tensor, so by $\ang{E,H,L}$-linearity and $\ov{V}(\ang{1}) = 1$ we have
\begin{itemize}
  \item (linear boundary) $\nu_{p,1} = p$.
\end{itemize}

These relations alone already determine the growth rate of the diagonal, by an elementary calculation.

\begin{lemma} \label{lem:nu-diagonal}
  Suppose that an array $\nu \colon \bbN^+ \times \bbN^+ \to \bbR$ satisfies
  \begin{itemize}
    \item (symmetry) $\nu_{p,q} = \nu_{q,p}$,
    \item (concavity) $\nu_{p,q} \geq (\nu_{p,q-1} + \nu_{p,q+1}) / 2$ for $q \geq 2$,
    \item (descent) $\nu_{p,3q+p-1} \geq 3\nu_{p,q}$, and
    \item (linear boundary) $\nu_{p,1} = p$.
  \end{itemize}
  Then we have the diagonal estimate $\nu_{p,p} \geq p^{4/3}$.
\end{lemma}
\begin{proof}
  Write $\Delta \nu_{p,q} = \nu_{p,q+1} - \nu_{p,q}$ for the difference in the second index. By concavity, $\Delta \nu_{p,q}$ is nonincreasing in $q$.

  Applying descent to $P_{p,q}$, we obtain $\nu_{p,3q+p-1} \geq 3 \nu_{p,q}$, which we rearrange as
  \[ 2\nu_{p,q} \leq \nu_{p,3q+p-1} - \nu_{p,q} = \sum_{r=q}^{3q+p-2} \Delta \nu_{p,r} \leq (2q+p-1) \, \Delta \nu_{p,q}. \]
  This gives
  \[ (2q+p+1) \, \nu_{p,q} \leq (2q+p-1) \, \nu_{p,q+1}. \]
  Since $\nu_{p,1} = p > 0$, induction on $q$ shows that $\nu_{p,q} > 0$ for all $p, q$, so after rearrangement we obtain
  \begin{equation}
    \frac{\nu_{p,q+1}}{\nu_{p,q}} \geq 1 + \frac{2}{2q+p-1}. \label{eqn:nu-ratio}
  \end{equation}

  By symmetry, $\nu_{p,p-1} = \nu_{p-1,p}$. Hence, applying \eqref{eqn:nu-ratio} with $(p, q) \to (p, p-1)$ and with $(p, q) \to (p-1, p-1)$, we get
  \[ \frac{\nu_{p,p}}{\nu_{p-1,p-1}} = \frac{\nu_{p,p}}{\nu_{p,p-1}} \cdot \frac{\nu_{p-1,p}}{\nu_{p-1,p-1}} \geq \left(1 + \frac{2}{3p-3}\right)\left(1 + \frac{2}{3p-4}\right) = \frac{(3p-1)(3p-2)}{(3p-3)(3p-4)}. \]
  Since $\nu_{1,1} = 1$ by the linear boundary condition, induction gives
  \[ \nu_{p,p} \geq \prod_{k=1}^{p-1} \left(1 + \frac{2}{3k}\right)\left(1 + \frac{2}{3k-1}\right). \]
  Finally, we have $1 + \frac{2}{3k-1} \geq 1 + \frac{2}{3k}$, and Bernoulli's inequality $(1+x)^{3/2} \geq 1 + \frac{3}{2}x$ with $x = \frac{2}{3k}$ gives $\left(1 + \frac{2}{3k}\right)^{3/2} \geq 1 + \frac{1}{k}$. Therefore,
  \[ \nu_{p,p} \geq \prod_{k=1}^{p-1} \left(1 + \frac{2}{3k}\right)^{2} \geq \prod_{k=1}^{p-1} \left(1 + \frac{1}{k}\right)^{4/3} = \left(\prod_{k=1}^{p-1} \frac{k+1}{k}\right)^{4/3} = p^{4/3}. \qedhere \]
\end{proof}

\begin{theorem} \label{thm:omega-nine-fourths}
  The matrix multiplication exponent $\omega$ is at most $9/4$.
\end{theorem}
\begin{proof}
  The tensor $P_{p,p}$ has minimal rank $2p-1$, so by \Cref{prop:catalytic-value} we have $\nu_{p,p} = \ov{V}(P_{p,p}) \leq \AR(P_{p,p})^{3/\omega} \leq (2p-1)^{3/\omega}$. On the other hand, $\nu$ satisfies all the hypotheses of \Cref{lem:nu-diagonal}, so $\nu_{p,p} \geq p^{4/3}$. Hence $p^{4/3} \leq (2p-1)^{3/\omega}$ for all $p \geq 1$. Taking logarithms and letting $p \to \infty$, we obtain $4/3 \leq 3/\omega$, that is, $\omega \leq 9/4$.
\end{proof}

\begin{remark}
  The original OpenAI preprint~\cite{OpenAI26NineFourths} states the bound only over $\bbC$. However, as we have seen, no step of the argument requires the field to have characteristic zero. The only minor technicality is that the claim that $P_{p,p}$ has rank $2p-1$ requires the ground field $\bbF$ to have sufficiently many elements. This is harmless: it has been a standard fact since Sch\"onhage~\cite[Thm.~2.8]{Sch81Tau} that $\omega$ depends only on the characteristic of the field, so we may assume without loss of generality that $\bbF$ is infinite.

  In fact, with sufficient care one can deduce the even stronger statement $\omega_\bbZ \leq 9/4$, where $\omega_\bbZ$ denotes the exponent of the rank of matrix multiplication tensors over $\bbZ$ (which in turn bounds the exponent of arithmetic circuits for matrix multiplication with coefficients in $\bbZ$). To see this, one replaces the rank decomposition of $P_{p,p}$ by a border rank decomposition over $\bbZ$, similar to the one in the footnote in the proof of \Cref{thm:semidirect-sum-restriction}, and reruns all the machinery of asymptotic rank, together with the above degenerations of polynomial multiplication tensors.
\end{remark}

\subsubsection{Spectral estimates} \label{subsubsec:spectral-estimates}

Let $\phi$ be a spectral point. We define the \emph{symmetrization} $\phi^\sym$ of $\phi$ as the geometric mean of the values of $\phi$ on all mode permutations of $T$:
\[ \phi^\sym(T) \coloneqq \phi \left( \bigotimes_{\sigma \in \mathfrak S_3} T^\sigma \right)^{1/6} = \prod_{\sigma \in \mathfrak S_3} \phi(T^\sigma)^{1/6}, \]
where the second equality holds since spectral points are multiplicative. Since $(T \otimes S)^\sigma = T^\sigma \otimes S^\sigma$, and $T \geq S$ implies $T^\sigma \geq S^\sigma$, the symmetrization $\phi^\sym$ is again multiplicative and monotone under restriction. Moreover, as $\sigma$ ranges over $\mathfrak S_3$, each of $E, H, L$ appears in each position of $\ang{E,H,L}^\sigma$ exactly twice, so
\[ \phi^\sym(\ang{E,H,L}) = (EHL)^\tau, \qquad \text{where } \tau = \frac{\theta_1 + \theta_2 + \theta_3}{3}. \]
It is then not hard to see that $\phi^\sym$ shares the other properties of the catalytic value.

\begin{proposition} \label{prop:spectral-sym}
  The symmetrization $\phi^\sym$ is superadditive and semi-superadditive. More precisely, with $\tau = (\theta_1 + \theta_2 + \theta_3) / 3$, we have
  \[ \phi^\sym(T \oplus S) \geq \phi^\sym(T) + \phi^\sym(S), \qquad \phi^\sym(T \boxplus S)^{1/\tau} \geq \phi^\sym(T)^{1/\tau} + \phi^\sym(S)^{1/\tau}, \]
  where the semidirect sum may share any of the three modes.
\end{proposition}
\begin{proof}
  Write $t_\sigma = \phi(T^\sigma)$ and $s_\sigma = \phi(S^\sigma)$. Since $(T \oplus S)^\sigma = T^\sigma \oplus S^\sigma$ and $\phi$ is additive, superadditivity amounts to
  \[ \prod_{\sigma \in \mathfrak{S}_3} (t_\sigma + s_\sigma)^{1/6} \geq \prod_{\sigma \in \mathfrak{S}_3} t_\sigma^{1/6} + \prod_{\sigma \in \mathfrak{S}_3} s_\sigma^{1/6}, \]
  which is Mahler's inequality, i.e., the superadditivity of the geometric mean.

  For semi-superadditivity, we run the same concentration argument as in the proof of \Cref{thm:value-semidirect}. Suppose that $T$ and $S$ share the $\varkappa$-th mode. For each $0 \leq k \leq n$, the tensor power $(T \boxplus S)^{\otimes n}$ restricts to a semidirect sum, sharing the $\varkappa$-th mode, of $\binom{n}{k}$ isomorphic copies of $T^{\otimes k} \otimes S^{\otimes (n-k)}$. By \Cref{thm:semidirect-sum-restriction} (applied after permuting the modes so that the shared mode becomes the third one), this semidirect sum asymptotically restricts to $M \otimes T^{\otimes k} \otimes S^{\otimes (n-k)}$, where $M$ is a matrix multiplication tensor with $EHL = \binom{n}{k}$. Since $\phi^\sym$ is multiplicative, monotone, and satisfies $\phi^\sym(M) = \binom{n}{k}^\tau$ regardless of which mode is shared, we obtain
  \[ \phi^\sym(T \boxplus S)^n \geq \binom{n}{k}^{\tau} \phi^\sym(T)^k \, \phi^\sym(S)^{n-k}. \]
  Taking $1/\tau$-th powers and summing over $k$, we get
  \[ (n+1) \, \phi^\sym(T \boxplus S)^{n/\tau} \geq \sum_{k=0}^n \binom{n}{k} \phi^\sym(T)^{k/\tau} \, \phi^\sym(S)^{(n-k)/\tau} = \bigl(\phi^\sym(T)^{1/\tau} + \phi^\sym(S)^{1/\tau}\bigr)^n. \]
  Taking $n$-th roots and letting $n \to \infty$ gives the claim.
\end{proof}

We now sketch how to reprove \Cref{thm:omega-nine-fourths} using spectral points. First, it is not hard to see that setting
\[ \nu_{p,q} = \phi^\sym(P_{p,q})^{1/\tau} \]
again yields an array $\nu$ satisfying the hypotheses of \Cref{lem:nu-diagonal}: symmetry holds since $\phi^\sym$ is invariant under permuting the modes, concavity and descent follow from \Cref{prop:P-split,prop:P-Q-split} together with \Cref{prop:spectral-sym}, and the linear boundary condition follows from $\phi^\sym(\ang{p,1,1}) = p^\tau$. Hence we again conclude that $\phi^\sym(P_{p,p}) \geq p^{4\tau/3}$.

On the other hand, we recall Strassen's \emph{duality theorem} for asymptotic rank (cf.~\cite[Thm.~3.2]{Str88AsymSpec}, \cite[\S3]{WZ26Spectra}).
\begin{theorem}[Strassen duality] \label{thm:strassen-duality}
  For every tensor $T$, we have
  \[ \AR(T) = \max_{\phi \in \calX} \phi(T). \]
\end{theorem}
In particular, since $\phi(\ang{n,n,n}) = n^{\theta_1 + \theta_2 + \theta_3}$, applying \Cref{thm:strassen-duality} to $\ang{n,n,n}$ shows that $\omega$ equals the maximum of $\theta_1 + \theta_2 + \theta_3 = 3\tau$ over all spectral points, and this maximum is attained.

\begin{proof}[Second proof of \Cref{thm:omega-nine-fourths}]
  Let $\phi$ be a spectral point attaining $\theta_1 + \theta_2 + \theta_3 = \omega$, so that $\tau = \omega/3$. Since every spectral point is bounded by the rank, and $\Rk(P_{p,p}^\sigma) = \Rk(P_{p,p})$ for every $\sigma \in \mathfrak S_3$, evaluating $\phi^\sym$ on $P_{p,p}$ gives
  \[ p^{4\tau/3} \leq \phi^\sym(P_{p,p}) \leq \Rk(P_{p,p}) \leq 2p-1. \]
  Letting $p \to \infty$, we get $4\tau/3 \leq 1$, that is, $\omega = 3\tau \leq 9/4$.
\end{proof}

\subsection{A closer look} \label{subsec:closer-look}

As we have seen, both types of functionals lead to the same system of inequalities for the array $\nu_{p,q}$, and hence to the diagonal estimates
\[ \ov{V}(P_{p,p}) \geq p^{4/3}, \qquad \phi^\sym(P_{p,p}) \geq p^{4\tau/3}. \]
We point out that each of these inequalities in fact shows that $P_{p,p}$, suitably symmetrized, asymptotically restricts to certain matrix multiplication tensors.

\paragraph{Spectral estimates.}
We begin by explaining what the estimate
\[ \phi^\sym(P_{p,p}) \geq p^{4\tau/3} \]
says. Unpacking the definition, we write the symmetrization of a tensor as
\[ T^\sym = \bigotimes_{\sigma \in \mathfrak{S}_3} T^\sigma, \]
so that $\phi^\sym(T) = \phi(T^\sym)^{1/6}$, and the estimate reads $\phi(P_{p,p}^\sym) \geq p^{8\tau}$. Recalling that $\tau = (\theta_1 + \theta_2 + \theta_3)/3$ and taking cubes, we obtain
\[ \phi\bigl((P_{p,p}^\sym)^{\otimes 3}\bigr) \geq p^{8(\theta_1 + \theta_2 + \theta_3)} = \phi(\ang{p^8, p^8, p^8}). \]
Since this holds for every spectral point $\phi$, it is in fact a statement about asymptotic restriction. This is the full version of Strassen duality (cf.~\cite[Thm.~3.2]{Str88AsymSpec}, \cite[\S3]{WZ26Spectra}), of which \Cref{thm:strassen-duality} only states the consequence for asymptotic rank.
\begin{theorem}[Strassen duality for asymptotic restriction] \label{thm:strassen-duality-restriction}
  For tensors $T$ and $S$, we have $T \gtrsim S$ if and only if
  \[ \phi(T) \geq \phi(S) \quad \text{for all } \phi \in \calX. \]
\end{theorem}
Hence \Cref{thm:strassen-duality-restriction} gives the asymptotic restriction
\[ (P_{p,p}^\sym)^{\otimes 3} \gtrsim \ang{p^8, p^8, p^8}. \]

In other words, $P_{p,p}^\sym$ asymptotically restricts to a square matrix multiplication tensor of volume $p^8$.

\paragraph{Value estimates.}
Does this say anything about $P_{p,p}$ itself, before symmetrization? For instance, does $P_{p,p}$ asymptotically restrict to a matrix multiplication tensor of volume $p^{4/3}$? This is not clear from the analysis above. At first glance, the catalytic value estimate
\[ \ov{V}(P_{p,p}) \geq p^{4/3} \]
does not directly help either: literally, it only provides catalytic degenerations $\ang{E_0, H_0, L_0} \otimes P_{p,p}^{\otimes n} \neged \ang{E_1, H_1, L_1}$ whose \emph{relative} volume $E_1 H_1 L_1 / (E_0 H_0 L_0)$ grows like $p^{(4/3-o(1))n}$. However, we now show that the catalyst, although it seems necessary at first glance, can in fact be removed.

\begin{proposition} \label{prop:remove-catalyst}
  Let $T \in \bbF^r \otimes \bbF^r \otimes \bbF^r$ be a tensor. If $\ov{V}(T) > r$, then the catalyst can be removed, in the sense that
  \[ \ov{V}(T) = \sup \Bigl\{ (EHL)^{1/n} : T^{\otimes n} \gtrsim \ang{E, H, L} \Bigr\}. \]
\end{proposition}
\begin{proof}
  We begin by analyzing an arbitrary catalytic degeneration
  \[ \ang{E_0, H_0, L_0} \otimes T^{\otimes n} \neged \ang{E_1, H_1, L_1}. \]
  In principle, we can eliminate the catalyst by the following argument. First, we continue the degeneration by a restriction:
  \[ \ang{E_0, H_0, L_0} \otimes T^{\otimes n} \neged \ang{E_1, H_1, L_1} \geq \ang{E_0, H_0, L_0} \otimes \ang{\lfloor E_1 / E_0 \rfloor, \lfloor H_1 / H_0 \rfloor, \lfloor L_1 / L_0 \rfloor}. \]
  By multiplicative cancellation (\Cref{lem:cancel}), we can remove the catalyst from both sides and obtain
  \[ T^{\otimes n} \gtrsim \ang{\lfloor E_1 / E_0 \rfloor, \lfloor H_1 / H_0 \rfloor, \lfloor L_1 / L_0 \rfloor}. \]
  However, this can be useless: one of the side lengths of this matrix multiplication tensor may be zero. We show that this cannot happen when $E_1 H_1 L_1 / (E_0 H_0 L_0) > r^n$, which, by the assumption $\ov{V}(T) > r$, is the case for degenerations whose contributions are close to $\ov{V}(T)$. To see this, we compare the flattening ranks of the first mode on both sides of the degeneration. Since the flattening ranks of $\ang{E,H,L}$ are $EH$, $HL$ and $LE$, we get
  \[ E_0 H_0 \cdot r^n \geq E_1 H_1, \]
  that is,
  \[ \frac{E_1 H_1}{E_0 H_0} \leq r^n < \frac{E_1 H_1 L_1}{E_0 H_0 L_0}, \]
  and we conclude that $L_1 > L_0$. Comparing the flattening ranks of the other two modes, we also conclude that $E_1 > E_0$ and $H_1 > H_0$.

  The rounding still loses a constant factor, but this loss can be amplified away by taking the $k$-th tensor power of the degeneration before removing the catalyst, which gives
  \[ T^{\otimes nk} \gtrsim \ang{\lfloor (E_1 / E_0)^k \rfloor, \lfloor (H_1 / H_0)^k \rfloor, \lfloor (L_1 / L_0)^k \rfloor}. \]
  Since the three ratios exceed $1$, the volume of the right-hand side is at least $\frac{1}{8} (E_1 H_1 L_1 / E_0 H_0 L_0)^k$, and its $(nk)$-th root tends to $(E_1 H_1 L_1 / E_0 H_0 L_0)^{1/n}$ as $k \to \infty$. Taking a family of catalytic degenerations whose contributions $(E_1 H_1 L_1 / E_0 H_0 L_0)^{1/n}$ tend to $\ov{V}(T)$, we conclude.
\end{proof}

Since the flattening ranks of $P_{p,p}$ are $p$, $p$ and $2p-1$, and $p^{4/3} > 2p-1$ for $p \geq 7$, \Cref{prop:remove-catalyst} applies to $P_{p,p}$ for all $p \geq 7$. Hence we can conclude that $P_{p,p}$ asymptotically restricts to matrix multiplication of volume $p^{4/3}$, in the sense that $P_{p,p}^{\otimes n} \gtrsim \ang{E,H,L}$ for some $E,H,L$ with $EHL \geq p^{(4/3 - o(1)) n}$.

\paragraph{Rectangular matrices.}
However, since the value does not distinguish the three side lengths $E, H, L$, we do not know exactly to which shape of matrix multiplication tensor $P_{p,p}$ asymptotically restricts. In principle, this can be read off from the spectral points. For reals $E, H, L \geq 1$, it essentially follows from \Cref{thm:strassen-duality-restriction} (applied to tensor powers, up to rounding) that $T^{\otimes n} \gtrsim \ang{E^{n-o(n)}, H^{n-o(n)}, L^{n-o(n)}}$ holds if and only if
\[ \phi(T) \geq E^{\theta_1} H^{\theta_2} L^{\theta_3} \quad \text{for all } \phi \in \calX. \]
To exploit this, we have to go back to the spectral values of the polynomial multiplication tensors without symmetrization. In principle, we should then keep track of three arrays rather than just one, corresponding to the three choices of which mode of the polynomial multiplication tensor serves as the output mode. The degenerations we constructed then lead to a more complicated family of inequalities relating these three arrays, since they relate a polynomial multiplication tensor to its mode permutations: for instance, the degeneration of \Cref{prop:P-Q-split} involves both $P_{p,q}$ and $Q_{p,q}$, the latter being $P_{p,q}$ with its last two modes swapped.

From this perspective, symmetrizing the spectral points is essentially a trick to simplify the analysis: it replaces the three exponents $\theta_1, \theta_2, \theta_3$ by their average $\tau = (\theta_1 + \theta_2 + \theta_3)/3$, so that we no longer need to track them separately. This trick works as long as we only care about the square exponent $\omega$. However, if we want bounds on the rectangular exponents $\omega(a,b,c)$, symmetrization loses too much information.

This asymmetric analysis has been carried out by Uzna\'nski~\cite{Uzn26Rect}.\footnote{The note~\cite{Uzn26Rect} was written by an AI model, and its terminology differs from the one commonly used in the literature (and in this note). We have restated its two main results in the standard language.} Its main result is the following estimate for $\specMM$. The reader may also consult the recent exposition of Amanov, van den Berg and Zuiddam~\cite{AdBZ26Poly}, which derives this estimate and its consequence \Cref{cor:rect-exponents} below in the language of the asymptotic spectrum~\cite[\S\S4.5--4.6]{AdBZ26Poly}, and moreover shows that the current ingredients cannot improve these bounds.

\begin{restatable}[{\cite[Prop.~5.2]{Uzn26Rect}; \cite[Prop.~4.14]{AdBZ26Poly}}]{theorem}{rectspectral} \label{thm:rect-spectral}
  Every $(\theta_1, \theta_2, \theta_3) \in \specMM$ satisfies
  \[ \theta_1 + \theta_2 + \theta_3 \leq 2 + 2\sqrt{(1 - \theta_1)(1 - \theta_2)(1 - \theta_3)}. \]
\end{restatable}

\begin{figure}[ht]
  \centering
  \includegraphics[width=\textwidth]{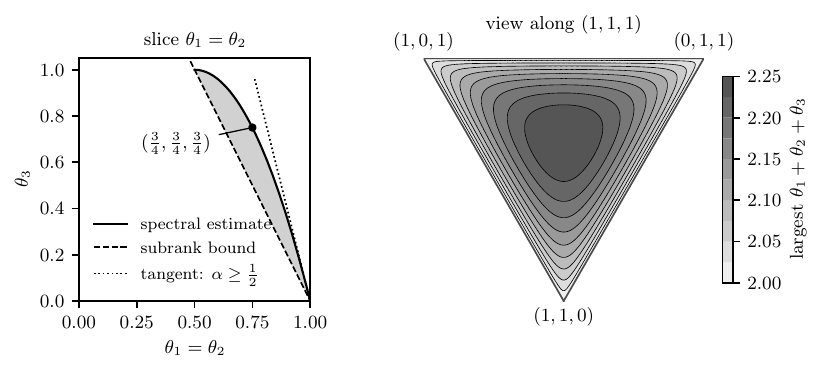}
  \caption{The region containing $\specMM$.}
  \label{fig:specMM}
\end{figure}

Together with the lower bound $\theta_1 + \theta_2 + \theta_3 \geq 2$, which follows from the subrank of matrix multiplication~\cite[Thm.~6.6]{Str87RelBilin}, \Cref{thm:rect-spectral} confines $\specMM$ to the thin region shown in \Cref{fig:specMM}.

Using the basic characterization
\[ \omega(a,b,c) = \max_{\theta \in \specMM} \, (a \theta_1 + b \theta_2 + c \theta_3), \]
which follows from \Cref{thm:strassen-duality} applied to $\ang{n^a, n^b, n^c}$, one obtains the following bounds for rectangular matrix multiplication.

\begin{restatable}[{\cite[Thm.~1.1]{Uzn26Rect}; \cite[Thm.~1.2]{AdBZ26Poly}}]{corollary}{rectexponents} \label{cor:rect-exponents}
  Let $a \geq b \geq c \geq 0$ with $b > 0$.
  \begin{enumerate}[label=(\roman*)]
    \item If $c \leq ab/(a+b)$, then $\omega(a,b,c) = a + b$.
    \item Otherwise, $\omega(a,b,c) \leq \dfrac{(ab + bc + ca)^2}{4abc}$.
  \end{enumerate}
\end{restatable}

For $a = b = c = 1$, case (ii) gives back $\omega \leq 9/4$. For $\omega(1,k,1)$, case (i) applies exactly when $k \leq 1/2$, and otherwise case (ii) gives $\omega(1,k,1) \leq 1 + k + \frac{1}{4k}$; in particular, the dual exponent satisfies $\alpha \geq 1/2$ and we also have $\omega(1,2,1) \leq 25/8$.

\section*{Background and acknowledgments}

\paragraph{Background.}
In the spring of 2026, while studying Strassen's support functionals~\cite{Str91SuppFunc} and the laser method, the author came up with the concept of semidirect sums, together with a conjecture formulated as \Cref{thm:semidirect-sum-restriction} and \Cref{thm:spectral-semidirect} of this note (it is not hard to see that these two statements are equivalent). One motivation for this conjecture is that it would improve various estimates in laser method analyses, e.g., it would give the hypothetical bound $\omega < 2.322$ discussed in \Cref{subsec:semidirect-sum}.

After testing the inequality of \Cref{thm:spectral-semidirect} on the support functionals, which by~\cite{SDW26Fun} are spectral points (and in fact the only explicitly known ones), the author found that it holds for them, and started to believe that it might hold in general. However, a proof remained out of reach despite several attempts, computational experiments, and consultations with AI models. It therefore came as a great surprise to the author that the technical lemmata of the recent OpenAI preprints~\cite{OpenAI26Below2258,OpenAI26NineFourths} settle exactly this conjecture.

In the author's view, although the new bounds on $\omega$ are significant and exciting, these lemmata are the main conceptual breakthrough of the OpenAI preprints: they provide new general methods for analyzing tensors, while the specific bounds can be seen as the outcome of a careful search for candidate tensors to which these methods apply. Since the attribution in the original preprints is poor, the author was motivated to write this exposition, in order to point out the techniques behind the original proofs of these lemmata, to discuss their relation to previous work, and to share his own thoughts about semidirect sums (most of \Cref{sec:intro} and part of \Cref{sec:untwisting}), which had crystallized before he learned of the proof of his conjecture.

\paragraph{Concurrent writings.}
After the OpenAI release, several groups of people have shared their own understanding of the new bound, each with a focus different from ours. The exposition of Amanov, van den Berg and Zuiddam~\cite{AdBZ26Poly} thoroughly works out the details of the proof, and shows that the analyses of the currently known degenerations of polynomial multiplication tensors are \emph{optimal}~\cite[\S5]{AdBZ26Poly}. The notes of Pass~\cite{Pas26Pack} and Shpilka~\cite{Shp26Expo} study the constructiveness of the $\omega \leq 9/4$ proof, and give quantitative bounds on the $o(1)$ term behind the $n^{9/4+o(1)}$ conclusion.

\paragraph{Acknowledgments.}
The author would like to thank Josh Alman and Kevin Pratt for inspiring discussions, for their continuous encouragement to think about semidirect sums and to write this exposition, and for kindly declining coauthorship. He also thanks them and J.~M.~Landsberg for valuable comments on an earlier draft of this note.

\paragraph{AI use.}
Claude Code was used for language editing, proofreading, checking computations, and producing \Cref{fig:specMM}. The author takes full responsibility for the correctness and presentation of the content.

\bibliographystyle{alphaurl}

\bibliography{main.bib}

\newcommand{\etalchar}[1]{$^{#1}$}
\begin{thebibliography}{{Ope}26b}

\bibitem[ADV{\etalchar{+}}25]{ADWXXZ25MoreAsym}
Josh Alman, Ran Duan, Virginia {Vassilevska Williams}, Yinzhan Xu, Zixuan Xu, and Renfei Zhou.
\newblock More asymmetry yields faster matrix multiplication.
\newblock In {\em Proceedings of the 2025 Annual ACM-SIAM Symposium on Discrete Algorithms ({SODA})}, pages 2005--2039, 2025.
\newblock \href {https://doi.org/10.1137/1.9781611978322.63} {\path{doi:10.1137/1.9781611978322.63}}.

\bibitem[AFL15]{AFL15Limit}
Andris Ambainis, Yuval Filmus, and Fran\c{c}ois {Le Gall}.
\newblock Fast matrix multiplication: Limitations of the {Coppersmith-Winograd} method.
\newblock In {\em Proceedings of the 47th Annual ACM Symposium on Theory of Computing ({STOC})}, pages 585--593, 2015.
\newblock \href {https://doi.org/10.1145/2746539.2746554} {\path{doi:10.1145/2746539.2746554}}.

\bibitem[AL26]{AL26Speedup}
Josh Alman and Baitian Li.
\newblock Asymptotic rank speedup theorems, revisited.
\newblock In {\em Proceedings of the 41st Computational Complexity Conference ({CCC})}, volume 383 of {\em Leibniz International Proceedings in Informatics ({LIPIcs})}, pages 36:1--36:42, 2026.
\newblock \href {https://doi.org/10.4230/LIPIcs.CCC.2026.36} {\path{doi:10.4230/LIPIcs.CCC.2026.36}}.

\bibitem[Alm21]{Alm21Barrier}
Josh Alman.
\newblock Limits on the universal method for matrix multiplication.
\newblock {\em Theory Comput.}, 17:Paper No. 1, 2021.
\newblock \href {https://doi.org/10.4086/toc.2021.v017a001} {\path{doi:10.4086/toc.2021.v017a001}}.

\bibitem[AV18a]{AW18Further}
Josh Alman and Virginia {Vassilevska Williams}.
\newblock Further limitations of the known approaches for matrix multiplication.
\newblock In {\em Proceedings of the 9th Innovations in Theoretical Computer Science Conference ({ITCS})}, volume~94 of {\em Leibniz International Proceedings in Informatics ({LIPIcs})}, pages 25:1--25:15, 2018.
\newblock \href {https://doi.org/10.4230/LIPIcs.ITCS.2018.25} {\path{doi:10.4230/LIPIcs.ITCS.2018.25}}.

\bibitem[AV18b]{AW18Limits}
Josh Alman and Virginia {Vassilevska Williams}.
\newblock Limits on all known (and some unknown) approaches to matrix multiplication.
\newblock In {\em Proceedings of the 59th Annual IEEE Symposium on Foundations of Computer Science ({FOCS})}, pages 580--591, 2018.
\newblock \href {https://doi.org/10.1109/FOCS.2018.00061} {\path{doi:10.1109/FOCS.2018.00061}}.

\bibitem[AV21]{AW21Refined}
Josh Alman and Virginia {Vassilevska Williams}.
\newblock A refined laser method and faster matrix multiplication.
\newblock In {\em Proceedings of the 2021 Annual ACM-SIAM Symposium on Discrete Algorithms ({SODA})}, pages 522--539, 2021.
\newblock \href {https://doi.org/10.1137/1.9781611976465.32} {\path{doi:10.1137/1.9781611976465.32}}.

\bibitem[AvdBZ26]{AdBZ26Poly}
Alimzhan Amanov, Maxim van~den Berg, and Jeroen Zuiddam.
\newblock Multiplying matrices via polynomials: Exposition and barriers, 2026.
\newblock \href {https://arxiv.org/abs/2610.30054} {\path{arXiv:2610.30054}}.

\bibitem[BCRL79]{BCLR79BRk}
Dario Bini, Milvio Capovani, Francesco Romani, and Grazia Lotti.
\newblock {$O(n^{2.7799})$} complexity for {$n \times n$} approximate matrix multiplication.
\newblock {\em Inf. Process. Lett.}, 8(5):234--235, 1979.
\newblock \href {https://doi.org/10.1016/0020-0190(79)90113-3} {\path{doi:10.1016/0020-0190(79)90113-3}}.

\bibitem[BCS97]{BCS97ACT}
Peter B{\"u}rgisser, Michael Clausen, and M.~Amin Shokrollahi.
\newblock {\em Algebraic Complexity Theory}, volume 315 of {\em Grundlehren der mathematischen Wissenschaften}.
\newblock Springer, Berlin, Heidelberg, 1997.
\newblock \href {https://doi.org/10.1007/978-3-662-03338-8} {\path{doi:10.1007/978-3-662-03338-8}}.

\bibitem[Beh46]{Beh46AP}
Felix~A. Behrend.
\newblock On sets of integers which contain no three terms in arithmetical progression.
\newblock {\em Proc. Natl. Acad. Sci. USA}, 32(12):331--332, 1946.
\newblock \href {https://doi.org/10.1073/pnas.32.12.331} {\path{doi:10.1073/pnas.32.12.331}}.

\bibitem[Bin80]{Bini80BRk}
Dario Bini.
\newblock Relations between exact and approximate bilinear algorithms. {A}pplications.
\newblock {\em Calcolo}, 17(1):87--97, 1980.
\newblock \href {https://doi.org/10.1007/BF02575865} {\path{doi:10.1007/BF02575865}}.

\bibitem[Bl{\"a}13]{Blaser13FMM}
Markus Bl{\"a}ser.
\newblock {\em Fast Matrix Multiplication}.
\newblock Number~5 in Theory of Computing Graduate Surveys. Theory of Computing Library, 2013.
\newblock \href {https://doi.org/10.4086/toc.gs.2013.005} {\path{doi:10.4086/toc.gs.2013.005}}.

\bibitem[BS23]{BS23AP}
Thomas~F. Bloom and Olof Sisask.
\newblock The {Kelley--Meka} bounds for sets free of three-term arithmetic progressions.
\newblock {\em Essent. Number Theory}, 2(1):15--44, 2023.
\newblock \href {https://doi.org/10.2140/ent.2023.2.15} {\path{doi:10.2140/ent.2023.2.15}}.

\bibitem[CFTZ22]{CFTZ22Corner}
Matthias Christandl, Omar Fawzi, Hoang Ta, and Jeroen Zuiddam.
\newblock Larger corner-free sets from combinatorial degenerations.
\newblock In {\em Proceedings of the 13th Innovations in Theoretical Computer Science Conference ({ITCS})}, volume 215 of {\em Leibniz International Proceedings in Informatics ({LIPIcs})}, pages 48:1--48:20, 2022.
\newblock \href {https://doi.org/10.4230/LIPIcs.ITCS.2022.48} {\path{doi:10.4230/LIPIcs.ITCS.2022.48}}.

\bibitem[CU03]{CU03Group}
Henry Cohn and Christopher Umans.
\newblock A group-theoretic approach to fast matrix multiplication.
\newblock In {\em Proceedings of the 44th Annual IEEE Symposium on Foundations of Computer Science ({FOCS})}, pages 438--449, 2003.
\newblock \href {https://doi.org/10.1109/SFCS.2003.1238217} {\path{doi:10.1109/SFCS.2003.1238217}}.

\bibitem[CU13]{CU13Coherent}
Henry Cohn and Christopher Umans.
\newblock Fast matrix multiplication using coherent configurations.
\newblock In {\em Proceedings of the 24th Annual ACM-SIAM Symposium on Discrete Algorithms ({SODA})}, pages 1074--1087, 2013.
\newblock \href {https://doi.org/10.1137/1.9781611973105.77} {\path{doi:10.1137/1.9781611973105.77}}.

\bibitem[CVZ21]{CVZ21Barrier}
Matthias Christandl, P\'eter Vrana, and Jeroen Zuiddam.
\newblock Barriers for fast matrix multiplication from irreversibility.
\newblock {\em Theory Comput.}, 17:Paper No. 2, 2021.
\newblock \href {https://doi.org/10.4086/toc.2021.v017a002} {\path{doi:10.4086/toc.2021.v017a002}}.

\bibitem[CVZ23]{CVZ23QFun}
Matthias Christandl, P\'eter Vrana, and Jeroen Zuiddam.
\newblock Universal points in the asymptotic spectrum of tensors.
\newblock {\em J. Amer. Math. Soc.}, 36(1):31--79, 2023.
\newblock \href {https://doi.org/10.1090/jams/996} {\path{doi:10.1090/jams/996}}.

\bibitem[CW90]{CW90Laser}
Don Coppersmith and Shmuel Winograd.
\newblock Matrix multiplication via arithmetic progressions.
\newblock {\em J. Symb. Comput.}, 9(3):251--280, 1990.
\newblock \href {https://doi.org/10.1016/S0747-7171(08)80013-2} {\path{doi:10.1016/S0747-7171(08)80013-2}}.

\bibitem[DEK{\etalchar{+}}26]{DEKMRSZAWB26AlphaEvolve}
Emilien Dupont, Marvin Eisenberger, Borislav Kozlovskii, Abbas Mehrabian, Francisco J.~R. Ruiz, Abigail See, Renfei Zhou, Josh Alman, Virginia {Vassilevska Williams}, and Matej Balog.
\newblock Improving the matrix multiplication exponent with modern optimization and {AlphaEvolve}, 2026.
\newblock \href {https://arxiv.org/abs/2608.16884} {\path{arXiv:2608.16884}}.

\bibitem[DS13]{DS13MM}
Alexander~M. Davie and Andrew~James Stothers.
\newblock Improved bound for complexity of matrix multiplication.
\newblock {\em Proc. Roy. Soc. Edinburgh Sect. A}, 143(2):351--369, 2013.
\newblock \href {https://doi.org/10.1017/S0308210511001648} {\path{doi:10.1017/S0308210511001648}}.

\bibitem[DWZ23]{DWZ23Hashing}
Ran Duan, Hongxun Wu, and Renfei Zhou.
\newblock Faster matrix multiplication via asymmetric hashing.
\newblock In {\em Proceedings of the 64th Annual IEEE Symposium on Foundations of Computer Science ({FOCS})}, pages 2129--2138, 2023.
\newblock \href {https://doi.org/10.1109/FOCS57990.2023.00130} {\path{doi:10.1109/FOCS57990.2023.00130}}.

\bibitem[KM23]{KM23AP}
Zander Kelley and Raghu Meka.
\newblock Strong bounds for 3-progressions.
\newblock In {\em Proceedings of the 64th Annual IEEE Symposium on Foundations of Computer Science ({FOCS})}, pages 933--973, 2023.
\newblock \href {https://doi.org/10.1109/FOCS57990.2023.00059} {\path{doi:10.1109/FOCS57990.2023.00059}}.

\bibitem[{Le }14]{LeGall14MM}
Fran\c{c}ois {Le Gall}.
\newblock Powers of tensors and fast matrix multiplication.
\newblock In {\em Proceedings of the 39th International Symposium on Symbolic and Algebraic Computation ({ISSAC})}, pages 296--303, 2014.
\newblock \href {https://doi.org/10.1145/2608628.2608664} {\path{doi:10.1145/2608628.2608664}}.

\bibitem[{Ope}26a]{OpenAI26NineFourths}
{OpenAI}.
\newblock {An Upper Bound of $9/4$ for the Matrix Multiplication Exponent}.
\newblock OpenAI Math Release preprint \href{https://github.com/openai/math/blob/main/preprints/Matrix-Multiplication-Nine-Fourths-October-2-2026/paper.pdf}{OAI:Matrix-Multiplication-Nine-Fourths-October-2-2026}, 2026.

\bibitem[{Ope}26b]{OpenAI26Below2258}
{OpenAI}.
\newblock {Complex Matrix Multiplication Below $2.258$ and Rectangular Bounds}.
\newblock OpenAI Math Release preprint \href{https://github.com/openai/math/blob/main/preprints/Complex-Matrix-Multiplication-Below-2.258-and-Rectangular-Bounds-September-24-2026/Complex-Matrix-Multiplication-Below-2.258-and-Rectangular-Bounds-September-24-2026.pdf}{OAI:Complex-Matrix-Multiplication-Below-2.258-and-Rectangular-Bounds-September-24-2026}, 2026.

\bibitem[{Ope}26c]{OpenAI26Staggered}
{OpenAI}.
\newblock {Staggered extraction for exact matrix multiplication over every field}.
\newblock OpenAI Math Release preprint \href{https://github.com/openai/math/blob/main/preprints/Staggered-extraction-for-exact-matrix-multiplication-over-every-field-September-24-2026/Staggered-extraction-for-exact-matrix-multiplication-over-every-field-September-24-2026.pdf}{OAI:Staggered-extraction-for-exact-matrix-multiplication-over-every-field-September-24-2026}, 2026.

\bibitem[Pas26]{Pas26Pack}
Rafael Pass.
\newblock Matrix multiplication from polynomial packings: A constructive and elementary proof of $\omega \leq 9/4$, 2026.
\newblock \href {https://arxiv.org/abs/2610.28627} {\path{arXiv:2610.28627}}.

\bibitem[Rag26]{Rag26AP}
Rushil Raghavan.
\newblock Improved bounds for 3-progressions, 2026.
\newblock \href {https://arxiv.org/abs/2603.27045} {\path{arXiv:2603.27045}}.

\bibitem[Rot53]{Rot53AP}
Klaus~F. Roth.
\newblock On certain sets of integers.
\newblock {\em J. Lond. Math. Soc.}, s1-28(1):104--109, 1953.
\newblock \href {https://doi.org/10.1112/jlms/s1-28.1.104} {\path{doi:10.1112/jlms/s1-28.1.104}}.

\bibitem[Sch81]{Sch81Tau}
A.~Sch\"onhage.
\newblock Partial and total matrix multiplication.
\newblock {\em SIAM J. Comput.}, 10(3):434--455, 1981.
\newblock \href {https://doi.org/10.1137/0210032} {\path{doi:10.1137/0210032}}.

\bibitem[SDW26]{SDW26Fun}
Keiya Sakabe, M.~Levent Do{\u{g}}an, and Michael Walter.
\newblock {Strassen's} support functionals coincide with the quantum functionals.
\newblock In {\em Proceedings of the 67th Annual IEEE Symposium on Foundations of Computer Science ({FOCS})}, 2026.
\newblock To appear.
\newblock \href {https://arxiv.org/abs/2601.21553} {\path{arXiv:2601.21553}}.

\bibitem[Shp26]{Shp26Expo}
Amir Shpilka.
\newblock Exposition of {OpenAI}'s matrix multiplication in exponent 9/4 (and an explicit algorithm).
\newblock Electronic Colloquium on Computational Complexity ({ECCC}), TR26-281, 2026.
\newblock URL: \url{https://eccc.weizmann.ac.il/report/2026/281/}.

\bibitem[Sto10]{Sto10Thesis}
Andrew~James Stothers.
\newblock {\em On the Complexity of Matrix Multiplication}.
\newblock PhD thesis, University of Edinburgh, 2010.
\newblock URL: \url{http://hdl.handle.net/1842/4734}.

\bibitem[Str69]{Str69Gaussian}
Volker Strassen.
\newblock Gaussian elimination is not optimal.
\newblock {\em Numer. Math.}, 13:354--356, 1969.
\newblock \href {https://doi.org/10.1007/BF02165411} {\path{doi:10.1007/BF02165411}}.

\bibitem[Str87]{Str87RelBilin}
V.~Strassen.
\newblock Relative bilinear complexity and matrix multiplication.
\newblock {\em J. Reine Angew. Math.}, 375/376:406--443, 1987.
\newblock \href {https://doi.org/10.1515/crll.1987.375-376.406} {\path{doi:10.1515/crll.1987.375-376.406}}.

\bibitem[Str88]{Str88AsymSpec}
V.~Strassen.
\newblock The asymptotic spectrum of tensors.
\newblock {\em J. Reine Angew. Math.}, 384:102--152, 1988.
\newblock \href {https://doi.org/10.1515/crll.1988.384.102} {\path{doi:10.1515/crll.1988.384.102}}.

\bibitem[Str91]{Str91SuppFunc}
V.~Strassen.
\newblock Degeneration and complexity of bilinear maps: Some asymptotic spectra.
\newblock {\em J. Reine Angew. Math.}, 413:127--180, 1991.
\newblock \href {https://doi.org/10.1515/crll.1991.413.127} {\path{doi:10.1515/crll.1991.413.127}}.

\bibitem[Uzn26]{Uzn26Rect}
Przemys{\l}aw Uzna{\'n}ski.
\newblock Rectangular matrix multiplication from polynomial products, 2026.
\newblock \href {https://arxiv.org/abs/2610.10494v2} {\path{arXiv:2610.10494v2}}.

\bibitem[{Vas}12]{Wil12MM}
Virginia {Vassilevska Williams}.
\newblock Multiplying matrices faster than {Coppersmith-Winograd}.
\newblock In {\em Proceedings of the 44th Annual ACM Symposium on Theory of Computing ({STOC})}, pages 887--898, 2012.
\newblock \href {https://doi.org/10.1145/2213977.2214056} {\path{doi:10.1145/2213977.2214056}}.

\bibitem[Vra24]{Vrana24Partial}
P\'eter Vrana.
\newblock Partial and weighted matrix multiplication, 2024.
\newblock \href {https://arxiv.org/abs/2408.15728} {\path{arXiv:2408.15728}}.

\bibitem[VXXZ24]{VXXZ24Rect}
Virginia {Vassilevska Williams}, Yinzhan Xu, Zixuan Xu, and Renfei Zhou.
\newblock New bounds for matrix multiplication: From alpha to omega.
\newblock In {\em Proceedings of the 2024 Annual ACM-SIAM Symposium on Discrete Algorithms ({SODA})}, pages 3792--3835, 2024.
\newblock \href {https://doi.org/10.1137/1.9781611977912.134} {\path{doi:10.1137/1.9781611977912.134}}.

\bibitem[WZ26]{WZ26Spectra}
Avi Wigderson and Jeroen Zuiddam.
\newblock Asymptotic spectra: Theory, applications and extensions.
\newblock {\em Bull. Amer. Math. Soc.}, 63(2):199--331, 2026.
\newblock \href {https://doi.org/10.1090/bull/1880} {\path{doi:10.1090/bull/1880}}.

\end{thebibliography}

\appendix

\section{Two laser methods} \label{app:two-laser-methods}

In this appendix, we discuss the intrinsic relation between the two kinds of laser methods: that of Strassen~\cite{Str87RelBilin} and that of Coppersmith and Winograd~\cite{CW90Laser}. We have long noticed that different communities in algebraic complexity theory tend to prefer different versions of the laser method, yet we have found no explicit discussion of this intrinsic connection in the literature, apart from a few brief remarks, e.g.,~\cite[Footnote~5]{CFTZ22Corner}; we therefore provide a thorough explanation here.

We begin with the setting of our presentation. Consider a tensor $T$ whose mode spaces are decomposed into subspaces $U = \bigoplus_{i \in I} U_i$, $V = \bigoplus_{j \in J} V_j$ and $W = \bigoplus_{k \in K} W_k$. Accordingly, we write $T$ as a sum of its components
\[ T = \sum_{i,j,k} T_{ijk}, \qquad T_{ijk} \in U_i \otimes V_j \otimes W_k, \]
and call the set $\Phi \coloneqq \{(i,j,k) : T_{ijk} \neq 0\} \subseteq I \times J \times K$ the \emph{blockwise support} of $T$.

\paragraph{Combinatorial framework.}
The goal of both laser methods is to find a combinatorial way to degenerate $T$ into a direct sum of some of the components $T_{ijk}$. Being combinatorial, such a way depends only on the underlying support structure $\Phi$.

The Coppersmith--Winograd laser method views $\Phi$ as a tripartite hypergraph, and seeks a large \emph{induced matching}. That is, we pick subsets $I' \subseteq I$, $J' \subseteq J$ and $K' \subseteq K$ of the same size $m$, such that the restriction $\Phi \cap (I' \times J' \times K')$ is a perfect matching: it consists exactly of $m$ triples $(i_t, j_t, k_t) \in \Phi$, $t = 1, \dots, m$, where $i_t, j_t, k_t$ run through $I', J', K'$ respectively. In this case, zeroing out all subspaces not indexed by $I', J', K'$ turns the tensor into a direct sum:
\[ T \geq \bigoplus_{t=1}^m T_{i_t j_t k_t}. \]

Strassen's laser method instead seeks a \emph{combinatorial degeneration}, which means finding \emph{gradings} $\alpha \colon I \to \bbZ$, $\beta \colon J \to \bbZ$ and $\gamma \colon K \to \bbZ$ such that
\begin{itemize}
  \item every $(i,j,k) \in \Phi$ satisfies $\alpha(i) + \beta(j) + \gamma(k) \geq 0$, and
  \item the triples $(i,j,k) \in \Phi$ attaining $\alpha(i) + \beta(j) + \gamma(k) = 0$ form a matching $\Phi_0$.
\end{itemize}
In this case, we can scale each $U_i$ by $\lambda^{\alpha(i)}$, and similarly each $V_j$ by $\lambda^{\beta(j)}$ and each $W_k$ by $\lambda^{\gamma(k)}$, where $\lambda$ is a formal parameter. This gives
\[ T \cong \sum_{i,j,k} \lambda^{\alpha(i) + \beta(j) + \gamma(k)} \, T_{ijk}, \]
and letting $\lambda \to 0$ yields the degeneration
\[ T \neged \bigoplus_{(i,j,k) \in \Phi_0} T_{ijk}. \]

\paragraph{Tight sets.}
It turns out that both laser methods find the situation easy when the support $\Phi$ is \emph{tight}, which means that one can find embeddings $I, J, K \subseteq \bbZ$ such that $i + j + k = 0$ for every $(i, j, k) \in \Phi$. Moreover, both methods find that one should reduce the general case to arguably the ``densest'' tight case\footnote{Actual applications of the laser method need to deal with certain sparse tight sets, which require a careful \emph{randomization} argument to reduce to the case we present here. Since this is technical and not related to our discussion, we omit the details.}: from now on, for a positive integer $n$, we let $\Phi_n$ denote the set
\[ \Phi_n \coloneqq \{ (i,j,k) : i + j = k, \ 0 \leq i,j,k < n \}. \]
It is not hard to see that $\Phi_n$ is the support of the truncated polynomial multiplication tensor $\ang{\bbF[t]/(t^n)}$. Here $\Phi_n$ is tight, since negating the third coordinate gives the desired equation. In what follows, we work with the equation $i + j = k$ directly.

We now demonstrate how the two laser methods obtain large matchings from $\Phi_n$.

\paragraph{Strassen's laser method.}
Strassen's laser method~\cite{Str87RelBilin} obtains a matching through a combinatorial degeneration. To be more precise, Strassen's original argument only treated the case where $\Phi$ is the support of a matrix multiplication tensor, which arises as the outer structure of the Strassen tensor presented in \Cref{subsec:semidirect-sum}. To our knowledge, the first presentations for general tight sets are due to the textbooks~\cite[\S15.8]{BCS97ACT} and~\cite[\S9]{Blaser13FMM}.

Since every triple of $\Phi_n$ already satisfies the linear equation $i + j = k$, a simple strategy to obtain a matching is to retain only those triples that satisfy a second linear equation. If this second hyperplane is not parallel to the first one, their intersection is a line, and with a suitable choice of the equation, the surviving triples form a matching. For simplicity, we take the line $i = j$.

This can be achieved by a monomial degeneration: we assign exponents to the coordinates in each mode so that the total exponent of each triple $(i, j, k) \in \Phi$ is exactly $(i - j)^2$. This is possible thanks to the factorization
\[ \lambda^{(i - j)^2} = \underbracket{\lambda^{2i^2}}_{U_i} \cdot \underbracket{\lambda^{2j^2}}_{V_j} \cdot \underbracket{\lambda^{-k^2}}_{W_k}, \qquad k = i + j, \]
which follows from $(i - j)^2 = 2i^2 + 2j^2 - (i + j)^2$. In other words, $\alpha(i) = 2i^2$, $\beta(j) = 2j^2$ and $\gamma(k) = -k^2$ form a combinatorial degeneration of $\Phi$: the total exponent $(i-j)^2$ is a square, so it is nonnegative and vanishes only when $i = j$, and the triples $(i, i, 2i)$ clearly form a matching. For $\Phi = \Phi_n$, this tells us that
\begin{center}
  $\Phi_n$ can be monomially degenerated to a matching of size $\lceil n/2 \rceil$.
\end{center}

It is not hard to see that the above argument works for any second affine linear equation, by sieving with its square; this is also the key to the proof of \Cref{lem:semidirect-sum-degeneration}.

\paragraph{Coppersmith--Winograd laser method.}
The Coppersmith--Winograd laser method~\cite{CW90Laser} obtains an induced matching, by observing that the problem is almost already solved by a known combinatorial construction, namely \emph{3-arithmetic-progression-free (3AP-free) sets}. We say that a subset $A \subseteq \bbZ$ is 3AP-free if every arithmetic progression inside it is trivial:
\[ x, y, z \in A, \ x + z = 2y \implies x = y = z. \]

One can clearly see that if $\Phi$ is the set of triples satisfying $i + k = 2j$, then a 3AP-free set $A$ can be directly applied to restrict all three coordinates, which yields an induced matching of size $\lvert A \rvert$. With a slight affine change of coordinates, the same works for any tight set; we show it for $\Phi_n$. If $A$ is a 3AP-free subset of $\{0, \dots, \lceil n/2 \rceil - 1\}$, then $\Phi_n$ has an induced matching of size $\lvert A \rvert$: we take
\[ I' = J' = A, \qquad K' = 2 \cdot A \coloneqq \{2a : a \in A\}. \]
Indeed, every surviving triple $(i, j, k)$ gives a solution $a + a' = 2a''$ with $i = a$, $j = a'$ and $k = 2a''$ for some $a, a', a'' \in A$. Since $A$ is 3AP-free, we must have $a = a' = a''$, so the surviving triples are exactly $(a, a, 2a)$ for $a \in A$, which form an induced matching.

It then remains to ask how large $A$ can be, and this question turns out to be deep. First, it is impossible to retain a linear number of triples, as the combinatorial degeneration does: Roth~\cite{Rot53AP} first proved that $\lvert A \rvert = o(n)$ must be sublinear. After a long line of research, it was recently proved~\cite{KM23AP,BS23AP,Rag26AP} that we must have $\lvert A \rvert \leq n / 2^{\Omega((\log n)^{\beta})}$ for some $\beta > 0$. On the other hand, Behrend~\cite{Beh46AP} gave a construction almost matching this type of bound.
\begin{theorem}[\cite{Beh46AP}] \label{thm:behrend}
  There is a 3AP-free subset of $\{0, \dots, n-1\}$ of size $n / 2^{O(\sqrt{\log n})}$.
\end{theorem}

Combining \Cref{thm:behrend} with the above reduction, we obtain the following.

\begin{corollary} \label{cor:cw-matching}
  The set $\Phi_n$ has an induced matching of size $n / 2^{O(\sqrt{\log n})}$.
\end{corollary}

\subsection{From degenerations to zeroing outs}
At first glance, there seems to be no relation between Strassen's combinatorial degeneration and Coppersmith and Winograd's use of 3AP-free sets. However, this is not the case.

In this subsection, we show a general way to turn a combinatorial degeneration into a zeroing out, at a cost that depends on the \emph{order} of the degeneration. Readers familiar with tensor degenerations may recall the analogous argument that turns a border rank estimate into a rank estimate. We remark that this connection has already been noted in~\cite[Thm.~4.11]{CVZ23QFun} and~\cite[Cor.~4.2 of the arXiv version]{AW18Limits}.

We first rephrase the combinatorial degeneration so that its order can be defined. Instead of requiring the minimum of $\alpha(i) + \beta(j) + \gamma(k)$ to be zero, we add constants to $\alpha$, $\beta$ and $\gamma$, so that we may assume that they take nonnegative values. Then the minimum of $\alpha(i) + \beta(j) + \gamma(k)$ over $\Phi$ is some integer $h \geq 0$, which we call the \emph{order} of the degeneration; the retained triples are exactly those with $\alpha(i) + \beta(j) + \gamma(k) = h$. In this case, we say that $\Phi$ combinatorially degenerates to the matching $\Phi_0$ with order $h$.

Looking back at the argument for Strassen's laser method, we have the following.
\begin{proposition} \label{prop:strassen-order}
  The set $\Phi_n$ combinatorially degenerates to a matching of size $\lceil n/2 \rceil$ with order $(n-1)^2$.
\end{proposition}
\begin{proof}
  We shift the gradings of Strassen's sieve to nonnegative values, by taking $\alpha(i) = 2i^2$, $\beta(j) = 2j^2$ and $\gamma(k) = (n-1)^2 - k^2$. Then the total exponent of each $(i,j,k) \in \Phi_n$ is $(i-j)^2 + (n-1)^2$, whose minimum $(n-1)^2$ is attained exactly by the $\lceil n/2 \rceil$ triples $(i, i, 2i)$.
\end{proof}

We now show how to turn a combinatorial degeneration directly into a zeroing out. The argument below is essentially the weight-layer argument of~\cite[Lem.~6]{AW18Further}, which states an asymptotic version with an unspecified constant; here we make the dependence on $h$ explicit.
\begin{lemma} \label{lem:degen-to-zeroing}
  If $\Phi$ combinatorially degenerates to a matching $\Phi_0$ of size $D$ with order $h$, then $\Phi$ contains an induced matching of size at least $D / (h+1)^2$.
\end{lemma}
\begin{proof}
  The argument is simply an averaging principle. Each retained triple $(i, j, k) \in \Phi_0$ has a \emph{degree triple} $(\alpha(i), \beta(j), \gamma(k))$, which consists of nonnegative integers summing to $h$. There are at most $(h+1)^2$ such degree triples, so some degree triple $(a, b, c)$ is shared by at least $D / (h+1)^2$ retained triples. We take
  \[ I' = \alpha^{-1}(a), \qquad J' = \beta^{-1}(b), \qquad K' = \gamma^{-1}(c), \]
  that is, we zero out all coordinates in the three modes whose degrees are not $a$, $b$ and $c$, respectively. Every triple $(i, j, k) \in \Phi \cap (I' \times J' \times K')$ satisfies $\alpha(i) + \beta(j) + \gamma(k) = a + b + c = h$, so it is a retained triple with degree triple $(a, b, c)$. Conversely, every such retained triple survives. Since $\Phi_0$ is a matching, these triples form an induced matching of the claimed size.
\end{proof}

\paragraph{From Strassen to Coppersmith--Winograd.}
For Strassen's laser method, this conversion is usually too expensive, since the order of the degeneration is large compared to $D$: for $\Phi_n$, the order is $(n-1)^2$, while $D$ is only $\lceil n/2 \rceil$. However, the high order becomes much less problematic once we take tensor powers. Indeed, the $N$-th power $\Phi^{\times N} \subseteq I^N \times J^N \times K^N$ is the blockwise support of $T^{\otimes N}$, and the gradings
\[ \alpha^{(N)}(i_1, \dots, i_N) \coloneqq \alpha(i_1) + \cdots + \alpha(i_N), \]
and similarly $\beta^{(N)}$ and $\gamma^{(N)}$, show that $\Phi^{\times N}$ combinatorially degenerates to the matching $\Phi_0^{\times N}$ of size $D^N$ with order $Nh$. By \Cref{lem:degen-to-zeroing}, we obtain an induced matching of size at least
\[ \frac{D^N}{(Nh+1)^2} = D^{N - o(N)}. \]
Thus, for sufficiently large $N$, we obtain a large induced matching by zeroing out coordinates, and asymptotically nothing is lost compared with the degeneration.

We now use this observation to show that the argument alone already recovers \Cref{cor:cw-matching}, without using \Cref{thm:behrend}.
\begin{proof}[Alternative proof of \Cref{cor:cw-matching}]
  Applying \Cref{lem:degen-to-zeroing} to the $N$-th power of the degeneration in \Cref{prop:strassen-order}, we see that $\Phi_n^{\times N}$ has an induced matching of size at least
  \[ \frac{\lceil n/2 \rceil^N}{(N(n-1)^2 + 1)^2} \geq \frac{(n/2)^N}{N^2 n^4}. \]
  It remains to view $\Phi_n^{\times N}$ inside a single set $\Phi_{n'}$. We encode a vector $x \in \{0, \dots, n-1\}^N$ as the integer
  \[ \iota(x) \coloneqq \sum_{s=1}^N x_s (2n-1)^{s-1} < n' \coloneqq (2n-1)^N. \]
  Since the sum of two digits is at most $2n - 2$, no carries occur, so for $i, j, k \in \{0, \dots, n-1\}^N$ we have $\iota(i) + \iota(j) = \iota(k)$ if and only if $i + j = k$. In other words, after zeroing out all coordinates of $\Phi_{n'}$ outside the image of $\iota$, we are left with exactly a copy of $\Phi_n^{\times N}$. Hence $\Phi_{n'}$ also has an induced matching of size at least
  \[ \frac{(n/2)^N}{N^2 n^4} \geq \frac{n'}{4^N N^2 n^4}, \]
  where we used $n' \leq (2n)^N$. Choosing $n \approx 2^{\sqrt{\log n'}}$ and $N \approx \sqrt{\log n'}$, the loss $4^N N^2 n^4$ is $2^{O(\sqrt{\log n'})}$, so $\Phi_{n'}$ has an induced matching of size $n' / 2^{O(\sqrt{\log n'})}$. For a general $n'$ that is not of the form $(2n-1)^N$, we restrict to the largest such value below it, which only loses a constant factor.
\end{proof}

We remark to unfamiliar readers that, if one tracks the above argument carefully, it in fact produces a 3AP-free subset of $\{0, \dots, n-1\}$ of size $n / 2^{O(\sqrt{\log n})}$, that is, it reproves \Cref{thm:behrend}. Moreover, if one expands the argument, one finds that it is almost exactly Behrend's construction~\cite{Beh46AP}! That said, although the laser methods of Strassen and of Coppersmith and Winograd look very different, they rest on the same mechanism: Strassen's quadratic sieve, once expanded over tensor powers, is Behrend's construction in disguise.

\end{document}